\documentclass{jfm_arxiv}

\usepackage{graphicx}
\usepackage{epstopdf,epsfig}
\usepackage{newtxtext}
\usepackage{newtxmath}
\usepackage{natbib}
\usepackage[caption=false]{subfig}
\newcommand{\RomanNumeralCaps}[1]
\linenumbers

\newcommand{\bu}{\boldsymbol{u}}
\newcommand{\bx}{\boldsymbol{x}}

\usepackage{hyperref}
\hypersetup{
    colorlinks = true,
    urlcolor   = blue,
    citecolor  = blue,
}

\usepackage{amsmath}
\usepackage{color}
\usepackage{comment}

\graphicspath{{./figs/}}

\title[Detachment and infinite Reynolds limit in 2D flow]{Boundary-layer detachment and infinite-Reynolds limit of two-dimensional wall-bounded flows}

\author[C.S. Campolina \& J.\ Bec]{Ciro S.\ Campolina\aff{1}
  \corresp{\email{ciro.campolina@sns.it}} and
  J\'er\'emie Bec\aff{2}}

\affiliation{\aff{1}Scuola Normale Superiore, Piazza dei Cavalieri 7, 56126 Pisa, Italy
	\aff{2}Universit\'{e} C\^{o}te d'Azur, CNRS, Inria, INPHYNI, 17 rue Julien Laupr\^etre, 06200 Nice, France}

\begin{document}
\maketitle

\begin{abstract}
	Understanding boundary-layer detachment and its consequences for the infinite-Reynolds-number limit remains a major challenge in fluid mechanics.
    Here, direct numerical simulations of incompressible flow in a two-dimensional periodic channel with no-slip walls are used to investigate this problem.
    Reynolds-number-dependent initial conditions are constructed with a boundary layer of Prandtl thickness that converge in the energy norm to an analytic, stationary Euler flow, while the associated Prandtl system develops a finite-time singularity.
    Highly resolved simulations show boundary-layer detachment and departures from Prandtl scaling, yet provide numerical evidence that the Navier--Stokes solutions continue to converge to the Euler flow over the simulated time interval.
    In particular, the squared energy norm of the velocity difference and the time-averaged energy dissipation both scale approximately as $\mbox{\textit{Re}}^{-1/2}$, consistently with established convergence criteria.
    Although the detached vortices become more numerous and intense as the Reynolds number increases, their characteristic size and occupied area appear to decrease.
    These observations suggest how boundary-layer detachment and the breakdown of the Prandtl description can coexist with convergence to a smooth Euler flow.
\end{abstract}

\begin{keywords}
%Authors should not enter keywords on the manuscript, as these must be chosen by the author during the online submission process and will then be added during the typesetting process (see \href{https://www.cambridge.org/core/journals/journal-of-fluid-mechanics/information/list-of-keywords}{Keyword PDF} for the full list).  Other classifications will be added at the same time.
\end{keywords}

\section{Introduction}
\label{sec:intro}

Understanding the nonlinear and multiscale nature of turbulence remains a major challenge in theoretical physics.
Turbulent flows of simple molecular fluids are commonly described by the incompressible Navier--Stokes equations, which, in dimensionless form, read
\begin{equation}\label{EQ:Navier_Stokes}
	\partial_t \bu + \bu \cdot \nabla \bu = - \nabla p + \frac{1}{\Rey}\Delta \bu, \qquad \nabla \cdot \bu = 0 \qquad \text{in} \ \mathcal{D}.
\end{equation}
Here, $\bu(\bx,t)$ is the velocity field, $p(\bx,t)$ is the scalar pressure, $t$ is time, and $\bx$ is the spatial coordinate in the $d$-dimensional domain $\mathcal{D} \subset \mathbb{R}^d$, with $d \geq 2$.
The incompressibility constraint is expressed mathematically by the divergence-free condition.

The Reynolds number is the dimensionless parameter $\Rey = UL/\nu$, where $U$ and $L$ are characteristic velocity and length scales of the flow, respectively, and $\nu$ is the kinematic viscosity.
It quantifies the relative importance of inertial and viscous effects.
As the Reynolds number increases, nonlinearities predominate, and the range of dynamically active scales can broaden, often leading to increasingly complex turbulent flow behaviour~\citep{frisch1995turbulence}.

In the formal infinite-Reynolds-number limit, $\Rey\to\infty$, the viscous term vanishes, yielding the incompressible Euler equations,
\begin{equation}\label{EQ:Euler}
	\partial_t \overline{\bu} + \overline{\bu} \cdot \nabla \overline{\bu} = - \nabla \overline{p}, \qquad \nabla \cdot \overline{\bu} = 0 \qquad \text{in} \ \mathcal{D}.
\end{equation}
The behaviour of the Navier--Stokes solutions in this limit, and their convergence to solutions of the Euler equations, can depend strongly on whether the domain $\mathcal{D}$ has solid boundaries or not.

Fluid flows away from physical walls are often modelled in domains without boundaries, such as the full Euclidean space, $\mathbb{R}^d$, with suitable conditions at infinity, and the torus, $\mathbb{T}^d$.
In these settings, the Navier--Stokes solutions are known to converge to the corresponding Euler solution in the infinite-Reynolds-number limit, provided that a sufficiently regular Euler solution exists~\citep{swann1971convergence,kato1972nonstationary}.
In particular, this is the case of two-dimensional Euler flows with sufficiently regular initial conditions, which exhibit global-in-time regularity~\citep{wolibner1933theoreme,kato1967classical}.

In three dimensions, however, only local-in-time regularity is guaranteed~\citep{kato1972nonstationary}, and strong Euler solutions may develop finite-time singularities~\citep{elgindi2021finite}, characterised by the blow-up of velocity gradients and the consequent loss of regularity~\citep{beale1984remarks}.
Beyond such singularities, the solution should be continued in a weak formulation.
However, weak solutions of the Euler equations are strongly non-unique~\citep{wiedemann2011existence}, raising the fundamental question of how to select physically meaningful solutions~\citep{de2010admissibility}.

Following Onsager's conjecture~\citep{onsager1949statistical,eyink2006onsager} for dissipative Euler flows, weak solutions of the Euler equations should have regularity not greater than H\"older with exponent $1/3$~\citep{isett2018proof}, otherwise the flow would conserve the kinetic energy~\citep{constantin1994onsager,eyink1994energy}.
Still, the question of how to select one among many dissipative solutions persists, and it has been conjectured that, in fact, the correct selection principle should be based on a stochastic formulation~\citep{thalabard2020butterfly,bernal2026spontaneous}, since randomness could emerge from deterministic systems in the infinite-Reynolds limit, a phenomenon called spontaneous stochasticity~\citep{eyink2026beyond}.

The presence of solid boundaries introduces additional physical and mathematical challenges in high-Reynolds-number flows.
The Navier--Stokes equations~\eqref{EQ:Navier_Stokes} are classically supplemented by the no-slip boundary condition,
\begin{equation}\label{EQ:no_slip}
	\bu = \mathbf{0} \qquad \text{on} \ \partial \mathcal{D},
\end{equation}
whereas the Euler equations~\eqref{EQ:Euler} satisfy only the impermeability, or no-penetration, condition,
\begin{equation}\label{EQ:no_penetration}
	\overline{\bu} \cdot \mathbf{n} = 0 \qquad \text{on} \ \partial \mathcal{D},
\end{equation}
where $\mathbf{n}$ denotes the outward unit normal vector to the boundary.

If the Euler flow also has zero tangential velocity at the boundary for all times, simple energy estimates can establish convergence of the Navier--Stokes solutions to the Euler solution in the infinite-Reynolds-number limit~\citep{kelliher2017observations}---see, for instance~\S\ref{sec:maths} below.
In general, however, the mismatch between the boundary conditions~\eqref{EQ:no_slip} and~\eqref{EQ:no_penetration} for the two systems generates sharp velocity gradients near the boundary that are not bounded \textit{a priori}.

Considerable attention in the mathematical literature has been devoted to the infinite-Reynolds-number limit in settings with boundaries where a sufficiently regular solution of the Euler equations exists~\citep{lopes2019fluids}.
Such settings include many flows of physical interest, for example, flow past immersed bodies for which a potential-flow description is available~\citep{landau1959fluid}.

Existing convergence results for Navier--Stokes solutions towards a strong Euler solution can broadly be divided into two classes, both under restrictive conditions.
The first establishes local-in-time convergence for solutions with analytic data near the boundary~\citep{sammartino1998zeroI,sammartino1998zeroII,maekawa2013solution,maekawa2014inviscid}. 
The second concerns flows subject to specific symmetries or structural constraints, including two-dimensional flows with circular symmetry~\citep{matsui1994example,bona2002zero,lopes2008vanishingA,lopes2008vanishingB}, plane-parallel flows~\citep{mazzucato2008vanishing,mazzucato2011boundary,gie2019vanishing}, and parallel-pipe flows~\citep{mazzucato2010vanishing,han2012boundary,gie2019vanishing}.

Alternatively, conditional statements have been established, providing insightful characterisations of the convergence in the infinite-Reynolds-number limit.
One of the most classical results is due to \cite{kato1984remarks}, who showed that the convergence of the Navier--Stokes solutions to a strong Euler solution holds if and only if the time-averaged energy dissipation rate within a boundary layer of thickness proportional to $\Rey^{-1}$ vanishes as $\Rey\to\infty$.
Thus, if convergence to the Euler solution fails, non-vanishing energy dissipation must persist within a thin layer adjacent to the boundary as $\Rey\to\infty$.

A substantial body of research has built on Kato's criterion, leading to several refinements and alternative formulations.
In particular, the full velocity gradient in Kato's original criterion can be replaced by tangential derivatives along the boundary~\citep{temam1997behavior,wang2001kato} or by the vorticity~\citep{kelliher2007kato}.

Further characterisations of the infinite-Reynolds-number limit have also been provided.
\cite{kelliher2008vanishing} showed that the Navier--Stokes solutions approach the strong Euler solution if and only if a vortex sheet forms at the boundary.
Moreover, this behaviour is equivalent to the vanishing of the time-averaged power associated with viscous shear stresses acting against the tangential velocity of the Euler flow at the boundary~\citep{bardos2013mathematics,kelliher2017observations,bardos2018remarks}.
Further details of these convergence criteria are provided in~\S\ref{sec:maths} and~\S\ref{sec:vorticity}.

Sufficient conditions have also been established for the Navier--Stokes solutions to converge to a weak solution of the Euler equations.
We recall that weak-strong uniqueness may fail in domains with boundaries~\citep{wiedemann2018weak}.
Consequently, even when a strong Euler solution exists, the Navier--Stokes solutions may converge to a different, weak Euler solution, a possible scenario that raises fundamental physical questions~\citep{eyink2025weak,quan2026weak}.
Relevant results include conditions based on the scaling of local second-order structure functions~\citep{drivas2019remarks} and on vorticity measures away from the boundary~\citep{constantin2019vorticity}.
For related results on anomalous energy dissipation in domains with boundaries \citep[see][]{drivas2018onsager}.

Despite all the above important advances, describing the infinite-Reynolds-number limit for general flows in the presence of solid boundaries remains an open problem.
The major obstacle raised by the mismatch of boundary conditions~\eqref{EQ:no_slip} and~\eqref{EQ:no_penetration} is the development of a boundary layer of sharp gradients in the vicinity of the boundary.

In his boundary-layer theory, \cite{prandtl1905uber} hypothesised that, in a flow past a solid wall, viscous effects are confined to a thin region near the boundary, while the flow in the bulk is effectively inviscid.
The thickness $\delta$ of this boundary layer is expected to scale with the Reynolds number as $\delta\sim \Rey^{-1/2}$, reflecting the balance between viscous diffusion in the wall-normal direction and convection along the boundary.

For two-dimensional flows, $\bu(x,y,t) = (u,v)$, in the upper half-plane, $y >0$, this asymptotic description suggests that the Navier--Stokes velocity field consists of an outer Euler flow away from the wall,
\begin{equation}\label{EQ:Prandtl_expansion_1}
	\bu(x,y,t) \approx \overline{\bu}(x,y,t) \quad \text{for} \ y \gg \delta,
\end{equation}
and a boundary-layer correction near the wall,
\begin{equation}\label{EQ:Prandtl_expansion_2}
	u(x,y,t) \approx u^P\left( x, \frac{y}{\delta},t \right), \quad v(x,y,t) \approx \delta v^P\left( x, \frac{y}{\delta},t \right) \quad \text{for} \ y \lesssim \delta,
\end{equation}
where $\bu^P(x,Y,t) = (u^P,v^P)$, with $Y = y/\delta$, satisfies the Prandtl boundary-layer equations
\begin{equation}\label{EQ:Prandtl}
	\partial_t u^P + u^P \partial_x u^P + v^P \partial_Y u^P = -\frac{d\overline{p}}{dx}\Big|_{y = 0} + \partial_Y^2 u^P, \quad \partial_x u^P + \partial_Y v^P = 0.
\end{equation}
The Prandtl equations~\eqref{EQ:Prandtl} are supplemented by the no-slip boundary conditions
\begin{equation}
	\bu^P = 0 \quad \text{at} \ Y = 0
\end{equation}
and the matching condition with the outer Euler flow,
\begin{equation}
	\lim_{Y \to \infty} u^P(x,Y,t) = \overline{u}(x,y,t)\big|_{y = 0}.
\end{equation}

As long as the Prandtl system is well-posed and the asymptotic expansion~\eqref{EQ:Prandtl_expansion_1}--\eqref{EQ:Prandtl_expansion_2} remains valid, this framework provides a description of the infinite-Reynolds-number limit in which the Navier--Stokes flow approaches the Euler flow away from the boundary.
However, rigorously justifying this asymptotic description and establishing convergence to the Euler solution are challenging mathematical problems~\citep{weinan2000boundary}.

In fact, the Prandtl equations~\eqref{EQ:Prandtl} are known to be well-posed only under very restrictive assumptions. 
These include analytic initial data~\citep{sammartino1998zeroI,lombardo2003well,kukavica2013local} and the case when the initial velocity profile is monotonic in the wall-normal variable and the outer Euler flow provides a favourable pressure gradient~\citep{oleinik1966mathematical,oleinik1999mathematical,xin2004global,alexandre2015well,masmoudi2015local}.

For initial data with more general regularity, the Prandtl system has been shown to be strongly ill-posed, both at the linear~\citep{gerard2010ill,gerard2012remarks} and nonlinear~\citep{guo2011note} levels.
In particular, \cite{gerard2012remarks} showed that for some infinitely smooth initial data, local-in-time solutions of the Prandtl system do not exist in the $H^1$ Sobolev class.
These ill-posedness results are obtained from a strong destabilization mechanism due to two ingredients: viscosity and critical points in the base velocity profile~\citep{gerard2010ill}.
This last ingredient highlights the importance of having monotonic data in the boundary-layer profile, a condition which may cease to hold when adverse pressure gradients develop.

Moreover, even when the Prandtl system is well-posed initially, its solutions may develop finite-time singularities.
This can occur both with zero~\citep{e1997blowup} and with certain non-trivial outer Euler flows~(\citealt{vandommelen1980spontaneous,kukavica2017vandommelen}; see also~\citealt{dalibard2019separation}) for singularities in the stationary Prandtl equation.
\citet{kukavica2017vandommelen} characterised the singularity by the blow-up of the displacement thickness, which quantifies the displacement of the inviscid outer flow induced by the boundary layer~\citep{schlichting2016boundary}.
Such behaviour indicates that the boundary-layer approximation may cease to describe a thin region confined near the wall, signalling the onset of boundary-layer separation, also referred to as detachment. 
This phenomenon has been conjectured to be associated with the breakdown of the Prandtl expansion~\citep{nguyen2018energy}.

\cite{prandtl1905uber} himself recognised that the validity of the boundary-layer assumptions could be limited, noting that the boundary layer may separate from the physical wall and significantly affect the outer flow~\citep{schlichting2016boundary}.
Following separation, the standard Prandtl expansion may no longer provide an adequate description of the infinite-Reynolds-number limit, despite being specifically derived as an asymptotic model for that limit~\citep{cowley2001laminar}.
Understanding the behaviour of Navier--Stokes solutions during boundary-layer detachment therefore remains a challenging problem~\citep{cassel2000comparison}, requiring mathematical tools that go beyond the classical Prandtl theory.

Given the scarcity of rigorous mathematical results on boundary-layer separation~\citep{vasseur2023boundary,vasseur2024layer}, considerable effort has been devoted to numerical investigations of this phenomenon~\citep{clercx2009two}.
These studies stressed the role of no-slip boundaries as effective sources of vorticity, and hence enstrophy, in the flow~\citep{clercx2005no,van2006effects,kramer2008large}.
This raises the possibility of a dissipation anomaly in the infinite-Reynolds-number limit, which, in the light of Kato's theorem~\citep{kato1984remarks}, would be associated with a failure of the Navier--Stokes solutions to converge to the strong Euler solution.
This scenario is also plausible in two-dimensional domains with boundaries.
This contrasts with the case of domains without physical boundaries, where the enstrophy remains uniformly bounded with respect to the Reynolds number, thereby precluding anomalous energy dissipation.

Enstrophy production at the boundary is particularly pronounced during boundary-layer detachment~\citep{kramer2007vorticity}.
A paradigmatic configuration used to investigate this phenomenon is the vortex dipole, consisting of two vortices of opposite sign, interacting with a rigid flat wall.
This configuration was first introduced by~\cite{orlandi1990vortex} and has since been studied by several authors for different purposes~\citep{clercx2017dissipation}.
Initially located away from the wall, the dipole travels towards it under the action of its self-induced velocity. 
Upon impact with the wall, boundary-layer separation occurs, generating vorticity that causes the dipole to rebound into the bulk.

The first systematic numerical studies of the dipole--wall collision reported a Prandtl-type scaling of the total enstrophy, $Z\sim \Rey^{1/2}$, at high Reynolds numbers~\citep{clercx2002dissipation,keetels2011reynolds}. If this scaling persists as $\Rey\to\infty$, it would rule out anomalous energy dissipation in that limit.
In a subsequent numerical investigation, \cite{nguyen2011energy} emphasised the importance of adequately resolving the Kato layer, whose thickness scales as $\mathcal{O}(\Rey^{-1})$.
Using a volume-penalisation method together with highly resolved spectral discretisations, they observed a scaling $Z\sim \Rey$ during boundary-layer detachment.
This scaling is consistent with a Reynolds-number-independent energy dissipation rate and therefore with the possibility of anomalous energy dissipation in the infinite-Reynolds-number limit.

Following a controversy over the imposition of boundary conditions in the volume-penalisation method~\citep{sutherland2013effect}, \cite{nguyen2018energy} revisited the dipole-wall collision using compact finite differences, which allow the no-slip boundary condition to be imposed exactly.
They employed highly refined meshes to resolve the Kato layer.
In addition to solving the Navier--Stokes equations, they numerically investigated the corresponding non-trivial Euler--Prandtl system.
They reported finite-time blow-up of the Prandtl equations coinciding with the onset of boundary-layer separation in their Navier--Stokes simulations.
They also found a consistent $Z\sim \Rey$ scaling during detachment, providing further support for the dissipation anomaly reported in their earlier study~\citep{nguyen2011energy}.
Most notably, the contrasting behaviour of the two systems---with Euler vortices continuing to slip along the wall while their Navier--Stokes counterparts rebound into the flow---provided strong numerical evidence against convergence to the Euler solution in a regime where the Prandtl system breaks down following boundary-layer detachment.

In the present work, we investigate the infinite-Reynolds-number limit of flows undergoing boundary-layer detachment.
We focus on the two-dimensional setting, in which globally regular solutions of the Euler equations are available, thereby avoiding complications associated with possible finite-time singularities in the inviscid flow.
This allows us to focus on the role of enstrophy injection through the boundary in the emergence of anomalous energy dissipation.

Inspired by the studies of~\cite{nguyen2011energy,nguyen2018energy}, we also consider a dipole of vortices, but in what we believe to be a more controllable configuration.
Rather than using an unsteady Lamb vortex, as in those studies, we construct a stationary solution of the Euler equations for which the associated Prandtl system is known to develop a finite-time singularity.
This choice provides an explicit expression for the Euler base flow, eliminating the need to solve the coupled Euler--Prandtl system numerically.

We initialise the Navier--Stokes simulations using this Euler flow augmented by a boundary-layer correction whose thickness follows the Prandtl scaling.
Starting from these carefully designed initial conditions, we solve the Navier--Stokes initial-value problem at increasing Reynolds numbers using spectral methods.
The numerical resolution is chosen to capture the relevant flow scales, including the possible formation of a Kato layer associated with anomalous energy dissipation.
Finally, we use established mathematical criteria to investigate the infinite-Reynolds-number behaviour of the resulting flows.

The remainder of the paper is organised as follows. 
In \S\ref{sec:model}, we introduce the flow configuration and describe the numerical methods employed in this study.
In \S\ref{sec:phenomenology}, we present the first observations from our simulations and discuss the physical phenomenology of the flow.
In \S\ref{sec:maths}, we apply several rigorous mathematical criteria to investigate the infinite-Reynolds-number limit.
In \S\ref{sec:vorticity}, we interpret the phenomenological observations and mathematical results through a detailed analysis of the vorticity distribution throughout the flow.
Finally, \S\ref{sec:conclusions} presents our concluding remarks.

\section{Model and methods}
\label{sec:model}

\begin{figure}
	\centering
	\includegraphics[width=0.9\columnwidth]{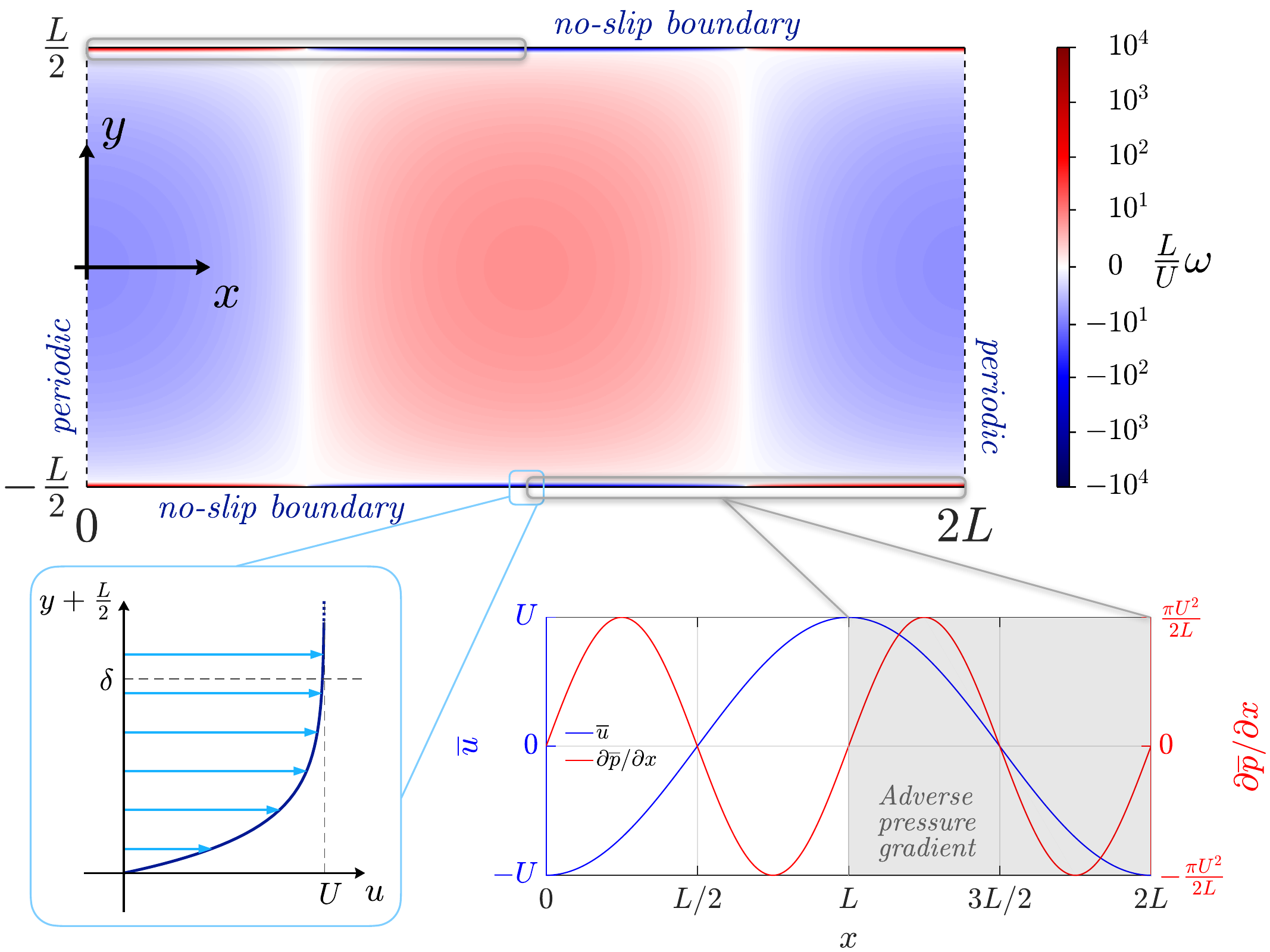}
	\caption{Flow configuration. \textit{Top:} Vorticity field corresponding to the Navier--Stokes initial condition in the two-dimensional periodic channel. The flow consists of two adjacent Taylor--Green vortex cells augmented by a boundary layer of thickness $\delta>0$. Colours are shown on a logarithmic scale using the continuous bi-symmetric logarithmic mapping of~\citet{webber2012bi}. \textit{Bottom,~left:}~detail of the boundary-layer velocity profile. \textit{Bottom,~right:}~Tangential velocities and tangential pressure gradient along the boundary for the Euler base flow. Regions of adverse pressure gradient are highlighted in grey, where boundary-layer separation is expected.}
	\label{fig:flow_configuration}
\end{figure}

In this section, we describe the flow configuration and the numerical methods employed to integrate the governing equations.

\subsection{Flow configuration}

The domain is the two-dimensional periodic channel
\begin{equation}
	\mathcal{D} = \mathbb{T} \times [-L/2,L/2],
\end{equation}
where $\mathbb{T} = \mathbb{R} / (2L \mathbb{Z})$ denotes the one-dimensional torus of length $2L$, identified with the interval $[0,2L]$ endowed with periodic boundary conditions.
Throughout, $\mathbf{x} = (x,y)$ represents the spatial coordinates in $\mathcal{D}$.
The boundary $\partial \mathcal{D}$ consists of the two lines $y = \pm L/2$.

We study the Cauchy problem for the incompressible Navier--Stokes equations subject to no-slip boundary conditions.
We consider $\Rey$-dependent initial data that converge to a prescribed Euler solution in the infinite-Reynolds limit, $\Rey \to \infty$.
The main idea is to consider a stationary solution of the Euler equations and, for the Navier–Stokes equations, to augment this Euler solution with a boundary layer in the initial data.

\subsubsection{Euler base flow}

The Euler base flow is defined as the restriction of a two-dimensional \citet{taylor1937mechanism} vortex to two adjacent cells:
\begin{equation}
	\overline{\bu}(x,y) =
	U
	\begingroup
	\renewcommand*{\arraystretch}{1.5}
	\begin{pmatrix}
		\phantom{-}\cos \frac{\pi x}{L} \sin \frac{\pi y}{L}\\
		-\ \sin \frac{\pi x}{L}   \cos \frac{\pi y}{L}
	\end{pmatrix}
	\endgroup
	, \quad (x,y) \in \mathcal{D}.
	\label{EQ:TG}
\end{equation}
Here, $U > 0$ denotes the characteristic velocity amplitude of the vortices.

The Taylor--Green vortex is a well-known stationary and analytic solution of the Euler equations on the two-dimensional torus.
When restricted to two adjacent cells as in~\eqref{EQ:TG}, this solution remains stationary and analytic on the two-dimensional periodic channel $\mathcal{D}$.
Moreover, it satisfies the no-penetration condition $\overline{\bu} \cdot \mathbf{n} = 0$ on $\partial \mathcal{D}$.
Other important features of this flow for our study will be discussed below in Sec.~\ref{SEC:features_flow}.

\subsubsection{Navier--Stokes initial condition}

We introduce a boundary layer of thickness $\delta>0$ into the Euler base flow and use the resulting velocity field as the initial condition for the Navier--Stokes equations.
To this end, we consider an analytic cut-off function $\varphi_\delta(y)$, depending only on the wall-normal coordinate $y$, satisfying
\begin{equation}
	\begin{cases}
		\varphi_\delta(y) = 0 \quad \text{if} \ y = \pm L/2\\
		\varphi_\delta(y) \simeq 1 \quad \text{if} \ \text{dist}(y,\partial \mathcal{D}) \gtrsim \delta.
	\end{cases}
\end{equation}
The Euler base flow augmented with a boundary layer of size $\delta$ is given by
\begin{equation}
	\bu^0(x,y) = \varphi_\delta(y)
	\overline{\bu}(x,y)
	+
	\phi_\delta(x,y)\hat{\mathbf{e}}_y
	, \quad (x,y) \in \mathcal{D},
\end{equation}
where $\hat{\mathbf{e}}_y = (0,1)$ is the unit vector in the $y$ direction, and $\phi_\delta(x,y)$ is an incompressibility correction given by
\begin{equation}
	\phi_\delta(x,y) = -\int_{0}^{y} \overline{v}(x,\tilde{y})\frac{\partial \varphi_\delta}{\partial y} (\tilde{y}) d \tilde{y},
\end{equation}
where $\overline{v}$ denotes the wall-normal component of $\overline{\mathbf u}$.

We now define the analytic cut-off function explicitly.
We consider the exponential boundary-layer profile
\begin{equation}
	\zeta(r) = 1 - e^{-r/c}.
\end{equation}
This smooth profile satisfies $\zeta(0) = 0$ and $\zeta(r) \to 1$ as $r \to \infty$, and the constant $c = 1/(2\log 10)$ is chosen so that $\zeta(1) = 0.99$, as is common in classical boundary-layer theory~\citep[see, e.g.,][]{schlichting2016boundary}.
For a boundary-layer thickness $\delta$, we define the rescaled boundary-layer profile
\begin{equation}
	\zeta_\delta(r) = \zeta(r/\delta),
\end{equation}
whose gradient concentrates in the region $0 \leq r \lesssim \delta$.
In particular, it satisfies $\zeta_\delta(\delta) = 0.99$.
The rescaled boundary-layer profile is shown in Fig.~\ref{fig:flow_configuration}.
Finally, we consider the analytic cut-off function
\begin{equation}
	\varphi_\delta(y) = \zeta_\delta(y+L/2)\zeta_\delta(L/2-y), \quad y \in [-L/2,L/2],
\end{equation}
which introduces identical boundary layers of thickness $\delta$ adjacent to the boundaries $y=\pm L/2$.
Fig.~\ref{fig:flow_configuration} shows a snapshot of the vorticity associated with the resulting velocity field $\bu^0$.

To make the Navier--Stokes initial condition Reynolds-number dependent, we adopt the Prandtl boundary-layer scaling for the boundary-layer thickness,
\begin{equation}
	\delta \sim \Rey^{-1/2}.
\end{equation}
This choice ensures that the initial boundary layer has a thickness consistent with classical \citet{prandtl1905uber} boundary-layer theory.

\subsubsection{Features of the flow configuration}
\label{SEC:features_flow}

This flow configuration has several theoretical and practical features that we summarise below.

The Euler base flow is an analytic stationary solution on $\mathcal{D}$ subject to the no-penetration boundary condition, $\overline{\bu} \cdot \mathbf{n} = 0$ on $\partial \mathcal{D}$.
Moreover, it exhibits non-zero slip at the boundary, given by
\begin{equation}\label{EQ:Euler_slip}
	\overline{\bu} \cdot \pmb{\tau} = - U \cos \frac{\pi x}{L} \quad \text{for} \ y = \pm \frac{L}{2},
\end{equation}
where $\pmb{\tau} = \mathbf{n}^\perp$ is the unit tangential vector along the boundary, and we adopt the convention $(x,y)^\perp = (-y,x)$.
As we shall see later, a non-zero slip at the boundary is a necessary condition for a possible lack of convergence of the Navier--Stokes solutions to the Euler strong solution~\citep{kelliher2017observations}---see \S\ref{sec:maths} below.

The tangential pressure gradient for the Euler base flow is
\begin{equation}\label{EQ:Euler_pressure_gradient}
	\nabla \overline{p} \cdot \pmb{\tau} = \mp \frac{\pi U^2}{2L} \sin \frac{2\pi x}{L} \quad \text{for} \ y = \pm \frac{L}{2}.
\end{equation}
From~\eqref{EQ:Euler_slip} and~\eqref{EQ:Euler_pressure_gradient}, we observe that adverse pressure gradients occur in the regions $ L \leq x \leq 2L$ along the bottom boundary $y = -L/2$ and $0 \leq x \leq L$ along the top boundary $y = L/2$, where the tangential velocity and the tangential pressure gradient have the same sign.
These are precisely the regions where boundary-layer separation may develop.
In Fig.~\ref{fig:flow_configuration}, we plot the tangential velocity and tangential pressure gradient along the bottom boundary and indicate the regions of adverse pressure gradient on the boundaries.

From a practical point of view, the Euler solution is available in explicit form and, therefore, there is no need to numerically integrate the incompressible Euler equations.

The Navier--Stokes initial conditions converge to the Euler base flow in energy norm in the infinite-Reynolds-number limit as
\begin{equation}
	\frac{1}{2}\int_\mathcal{D} | \bu^0(\mathbf{x}) - \overline{\bu}(\mathbf{x})|^2 \, d\mathbf{x} \sim \Rey^{-1/2} \to 0, \quad \text{when} \ \Rey \to \infty.
\end{equation}
The rate of convergence $\Rey^{-1/2}$ is precisely the scaling for the convergence of Prandtl boundary layer to the corresponding Euler outer flow.

The Navier--Stokes initial conditions are analytic in $\mathcal{D}$.
Therefore, for each $\Rey$ there exists a unique global-in-time smooth solution of the Navier--Stokes equations~\citep{temam1984navier}.
More importantly, the Prandtl boundary-layer equations are well-posed locally in time, and the Navier--Stokes solutions converge to the Euler strong solution on the corresponding finite-time interval~\citep{sammartino1998zeroI,sammartino1998zeroII}.
Nevertheless, the Euler base flow is chosen so that the boundary traces~\eqref{EQ:Euler_slip} and~\eqref{EQ:Euler_pressure_gradient} give rise to \citet{vandommelen1980spontaneous} singularity in the Prandtl equations, as rigorously established by~\cite{kukavica2017vandommelen}.

Physically, this means that the boundary layer initially remains attached (there is no boundary-layer separation), and that the Prandtl boundary-layer expansion holds true during the early stages of the evolution.
In this regime, the Navier--Stokes solutions converge to the Euler strong solution in the infinite-Reynolds limit.
However, this does not persist for all times, since the Prandtl boundary layer equations develop a finite-time singularity, beyond which the convergence of Navier--Stokes solutions to the Euler solution is no longer guaranteed.

\subsection{Numerical methods}

The incompressible Navier--Stokes equations are numerically integrated using Channelflow~2.0, an open-source software package for numerical simulations and analysis of the incompressible fluid flow in channel geometries~\citep[see][]{gibson2019channelflow2}.
The code is written in C++ and parallelised using the Message Passing Interface (MPI).

Channelflow 2.0 employs a spectral Galerkin method, with a standard Fourier expansion in the periodic variables and a Chebyshev expansion in the wall-normal direction $y$~\citep{boyd2001chebyshev}.
Computations are fully dealiased using \citet{orszag1971elimination} two-thirds rule.
Equations are solved in primitive variables, namely velocity and pressure~\citep[Sec.~7.3.1]{canuto1988spectral}, with the nonlinear convective term expressed in rotational form, $\bu \cdot \nabla \bu = (\nabla \times \bu) \times \bu + \frac{1}{2}\nabla ( \bu \cdot \bu )$.
This expression is mathematically equivalent to other common formulations of the nonlinear term when the incompressibility constraint, $\nabla \cdot \bu = 0$, is satisfied. 
In the discrete setting, however, the rotational form is computationally less expensive while providing the same numerical accuracy, as long as the highest spatial frequencies are properly dealiased~\citep{zang1991rotation}.
No-slip boundary conditions and the incompressibility constraint are enforced through the Influence Matrix Method~\citep{kleiser1980treatment}.

Time integration is performed using the 3rd-order semi-implicit backward differentiation formulae~(SBDF3)~\citep{ascher1995implicit}, a multi-step scheme that requires two initialization steps. These are solved with a three-substep, 2nd-order semi-implicit Runge--Kutta method (SMRK2) developed by~\cite{spalart1991spectral}.
All simulations are performed with a fixed time step, $\Delta t$, chosen such that the Courant--Friedrichs--Lewy (CFL) number remains below $0.6$ throughout each simulation.

Channelflow 2.0 is natively designed for flows on three-dimensional channels.
To perform two-dimensional simulations, we solved the three-dimensional governing equations using homogeneous initial data in the additional periodic $z$ direction while retaining only three Fourier modes in that direction.
This is the minimum number of modes compatible with the two-thirds dealiasing rule, resulting in two effective modes.
For every simulation, we verified that the solution remained identical in these two modes at all times, thereby ensuring that the computed flow was effectively two-dimensional.

\begin{table}
	\begin{center}
		\def~{\hphantom{0}}
		\begin{tabular}{cccccc}
			\textbf{Run}  & $N_x$   &   $N_y$ & $\Rey$ & $\delta$ & $\Delta t$ \\ \hline
			\textbf{I}   & 768 & 289  & 12 900 & $7.45 \times 10^{-3}$ & $1.27 \times 10^{-4}$\\
			\textbf{II}   & 1 152 & 433 & 22 000 & $5.70 \times 10^{-3}$ & $6.37 \times 10^{-5}$\\
			\textbf{III}  & 1 536 & 577 & 31 400 & $4.77 \times 10^{-3}$ & $6.37 \times 10^{-5}$\\
			\textbf{IV}   & 2 304 & 865 & 56 500 & $3.56 \times 10^{-3}$ & $3.18 \times 10^{-5}$\\
			\textbf{V} & 3 072 & 1 153 & 75 400 & $3.09 \times 10^{-3}$ & $3.18 \times 10^{-5}$\\
			\textbf{VI} & 4 608 & 1 729 & 138 000 & $2.28 \times 10^{-3}$ & $1.59 \times 10^{-5}$\\
			\textbf{VII} & 6 144 & 2 305 & 204 000 & $1.87 \times 10^{-3}$ & $1.59 \times 10^{-5}$\\
			~\textbf{VIII}~ & ~9 216~ & ~3 457~ & ~346 000~ & ~$1.44 \times 10^{-3}$~ & ~$6.37 \times 10^{-6}$~\\
		\end{tabular}
		\caption{Numerical parameters for each simulation run. $N_x$ and $N_y$ denote the number of Fourier and Chebyshev points, respectively, used in the spectral discretisation. With the two-thirds dealiasing rule, the numbers of retained modes are $2N_x/3$ and $2(N_y-1)/3+1$, respectively. $\Rey$ denotes the Reynolds number, $\delta$ is the initial boundary-layer thickness, and $\Delta t$ is the time step. Both $\delta$ and $\Delta t$ are given in dimensionless units.}
		\label{tab:runs}
	\end{center}
\end{table}

The numerical parameters used for each simulation run are summarised in Tab.~\ref{tab:runs}.
The number of Fourier modes, $N_x$, and Chebyshev polynomials, $N_y$, used in the spectral discretisation were chosen to ensure adequate resolution of the relevant flow scales.
To this end, the Reynolds number $\Rey$ was selected in relation to the maximum resolved wavenumber $k_\mathrm{max} = N_x / 3$ according to
\begin{equation}
	\Rey \approx 1.71 \, k_\mathrm{max}^{4/3}.
\end{equation}
As we will verify later, this choice provides a conservative resolution criterion for capturing the turbulent scales down to the smallest Kraichnan dissipation scale associated with the direct enstrophy cascade.

The number of Chebyshev polynomials was chosen to be proportional to $N_x$, yielding $N_y\sim \Rey^{3/4}$.
This resolution increases sufficiently rapidly with $\Rey$ to capture the Prandtl boundary-layer thickness, which scales as $\Rey^{-1/2}$.
Moreover, because Chebyshev points cluster quadratically near the boundaries, the local grid spacing there scales as $N_y^{-2}\sim \Rey^{-3/2}$.
This is asymptotically smaller than the thickness of a Kato layer, which scales as $\Rey^{-1}$~\citep{kato1984remarks}.
We will investigate the behaviour of the flow at this scale in~\S\ref{sec:maths}.

The initial boundary-layer thickness, $\delta$, was chosen according to the Prandtl boundary-layer scaling,
\begin{equation}
	\delta \approx 0.85 \, \Rey^{-1/2}.
\end{equation}

Finally, the time step was chosen to scale inversely with $N_x$:
\begin{equation}
	\Delta t \approx 0.1 \, N_x^{-1},
\end{equation}
resulting in a CFL number below 0.6 throughout each simulation.

Together, these choices were designed to provide well-resolved and conservative numerical simulations.
Further aspects of the numerical resolution and its adequacy are discussed alongside the results.

\section{First observations and phenomenology}
\label{sec:phenomenology}

\begin{figure}
	\centering
	\includegraphics[width=.8\columnwidth]{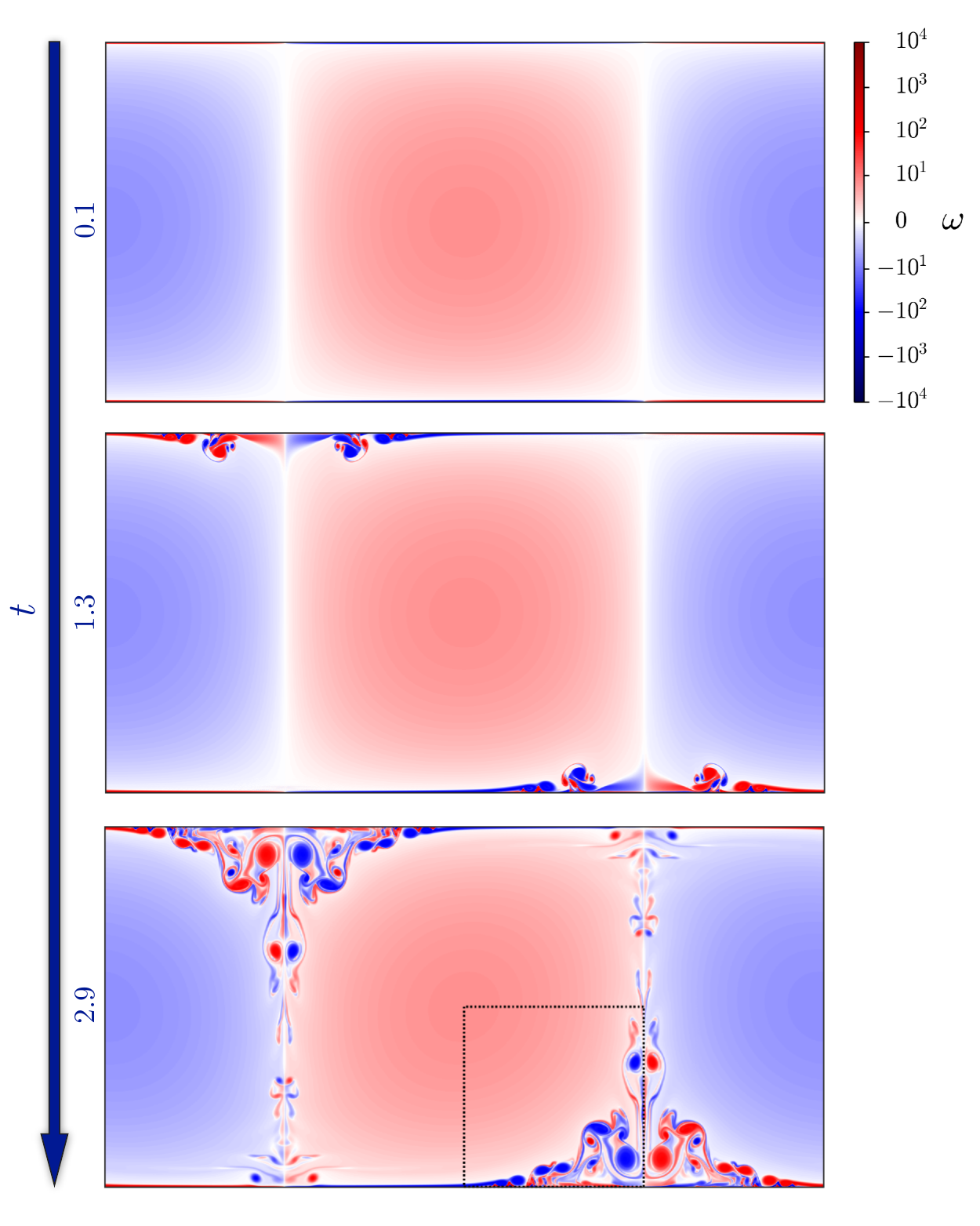}
	\caption{Vorticity snapshots at different times $t$ for Run~VI ($\Rey=138~000$). Colours are shown on the same logarithmic scale as in Fig.~\ref{fig:flow_configuration}.}
	\label{fig:vorticity_snapshots_domain}
\end{figure}

Figure~\ref{fig:vorticity_snapshots_domain} shows vorticity snapshots at different times $t$ for Run~VI, whose Reynolds number is $\Rey = 138~000$.
We observe that at the first instants of time ($t = 0.1$), the boundary layer remains attached, while the bulk flow is essentially unaffected by viscous effects.
In fact, the vorticity field is qualitatively indistinguishable from the initial condition displayed in Fig.~\ref{fig:flow_configuration}.

At a later time ($t = 1.3$), detachments occur at the expected regions of adverse pressure gradients, and the boundary layer separates from the wall.
As time evolves ($t = 2.9$), successive detachments develop and give rise to complex nonlinear interactions in a turbulent regime.
The detached vortices are advected by the large-scale flow and start to affect regions away from the boundary.

\begin{figure}
	\centering
	\includegraphics[width=\columnwidth]{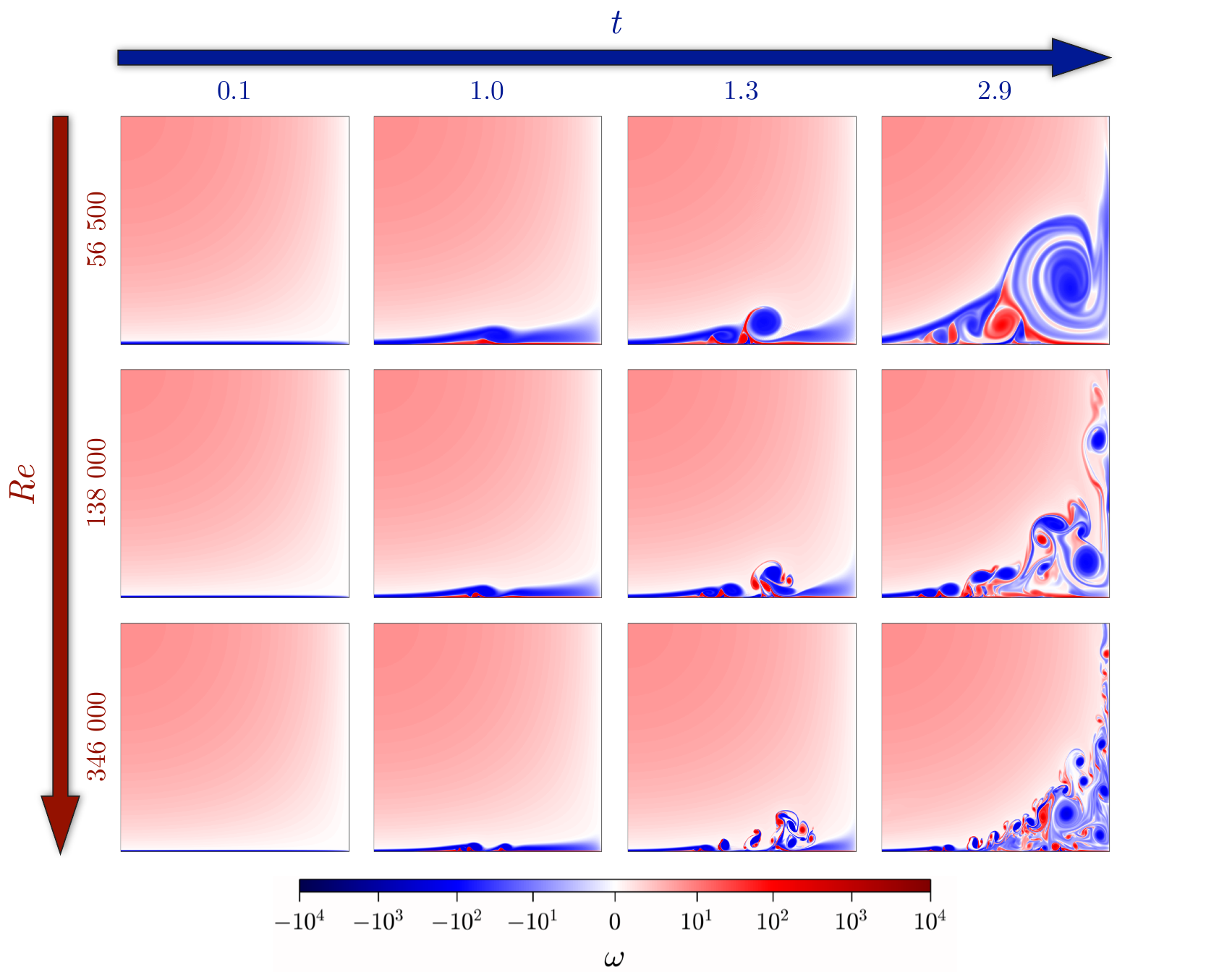}
	\caption{Close-up views of the vorticity field near the boundary at different times $t$ and Reynolds numbers $\Rey$. The region shown is outlined by a dashed black line in the bottom snapshot of Fig.~\ref{fig:vorticity_snapshots_domain}. All snapshots use the same logarithmic colour scale as in Fig.~\ref{fig:flow_configuration}.}
	\label{fig:vorticity_snapshots}
\end{figure}

Figure~\ref{fig:vorticity_snapshots} shows details of the vorticity field near the boundary at different times $t$ and for increasing Reynolds numbers.
From the snapshots, we observe that the boundary layer remains attached during the early stages of the evolution ($t=0.1$) for all Reynolds numbers, while the first detachments occur at approximately the same time ($t=1.0$).
However, as the Reynolds number increases, the detached vortices become smaller in size and more numerous, resulting in a more complex and intense flow regime.
More subtly, since all snapshots are shown using the same colour scale, we can also observe that the vorticity magnitude of the detached vortices increases with the Reynolds number.
This increase is particularly noticeable at the cores of the detached vortices.

We now investigate these events in greater detail through quantitative measurements of the relevant flow quantities.
We first relate the onset of boundary-layer separation to the breakdown of the Prandtl boundary-layer expansion.
We then examine the turbulent dynamics associated with the successive detachments that follow.

\subsection{Breakdown of the Prandtl expansion and onset of boundary-layer separation}

At early times, the boundary layer remains attached.
This regime corresponds to the short time interval over which the Prandtl boundary-layer equations are well-posed, and during which the Navier--Stokes solution is well approximated by the Prandtl expansion near the boundary and by the strong Euler solution in the bulk.
In this setting, convergence of the Navier--Stokes solutions to the strong Euler solution is therefore ensured~\citep{sammartino1998zeroII}.

We recall from classical boundary-layer theory~\citep{schlichting2016boundary} that, as long as the boundary layer remains attached, viscous effects are confined to a region of thickness $\delta \sim \Rey^{-1/2}$. Consequently, the following scalings are expected to hold:
\begin{equation}
	\omega |_{\partial \mathcal{D}} \sim \Rey^{1/2} \qquad \text{and} \qquad (\partial p /\partial y)|_{\partial \mathcal{D}} = \mathcal{O}(\Rey^{-1/2}).
\end{equation}

\begin{figure}
	\centering
	\subfloat{\includegraphics[width=0.45\columnwidth]{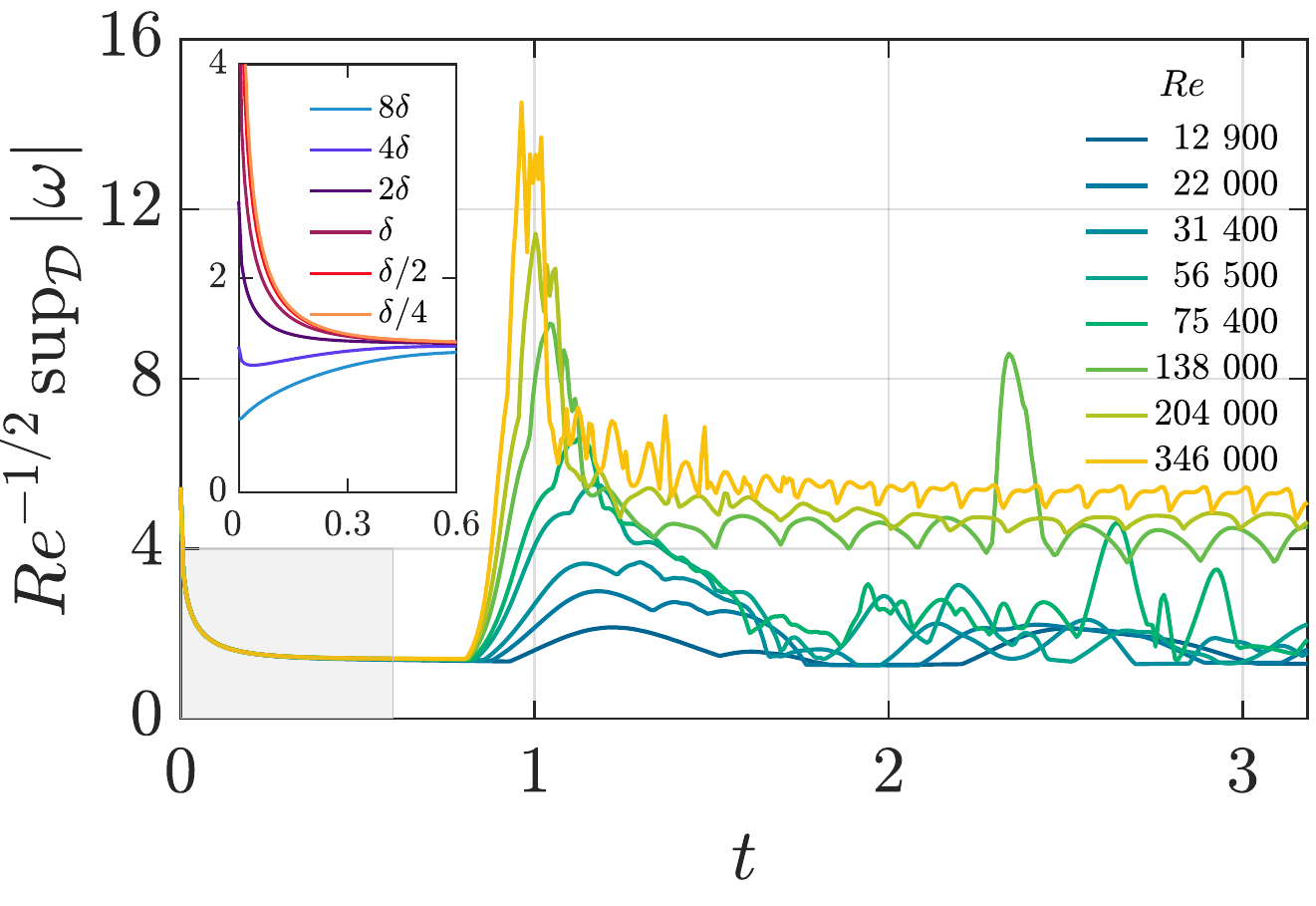}\label{fig:max_vorticity}}
	\qquad
	\subfloat{\includegraphics[width=0.45\columnwidth]{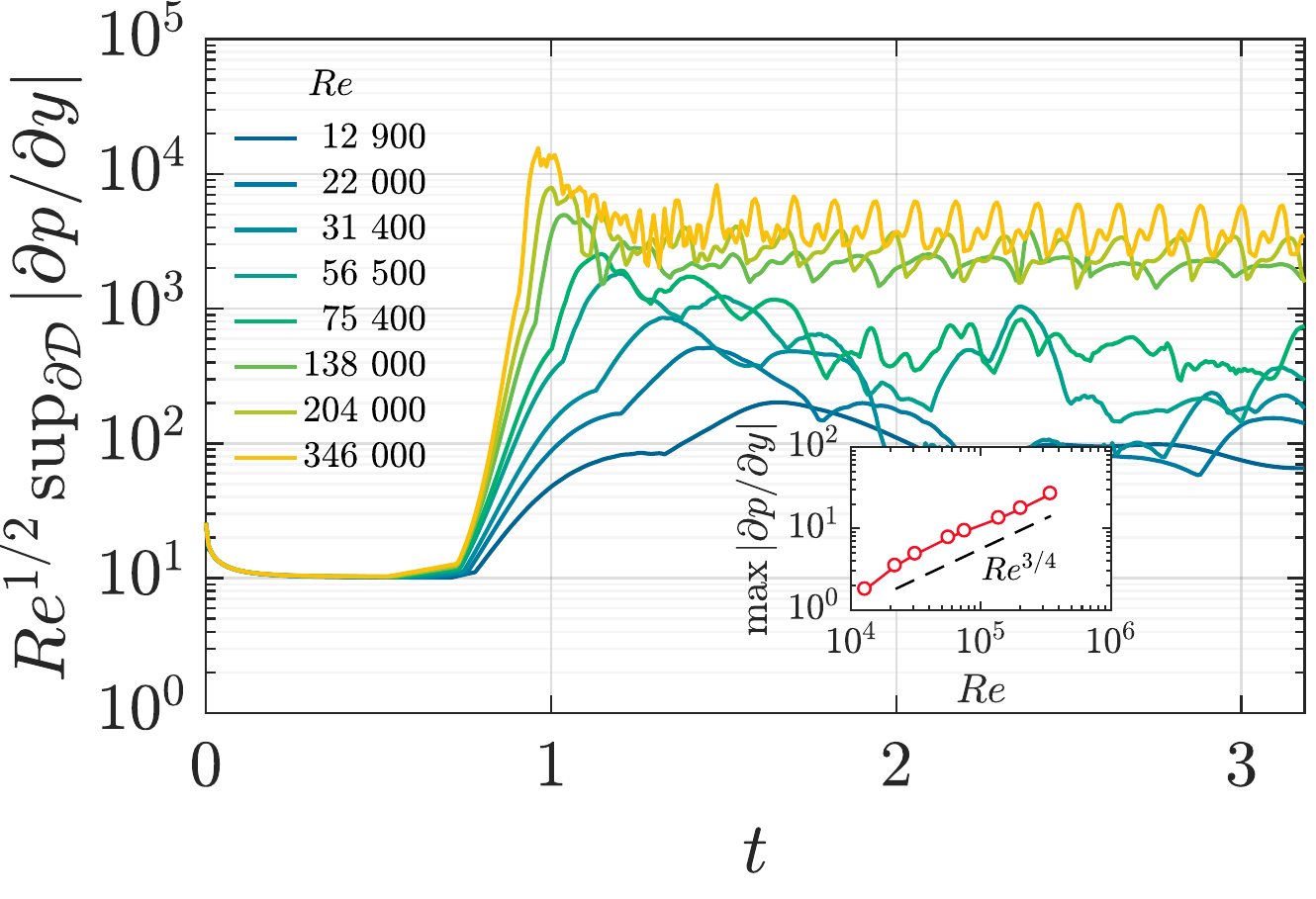}\label{fig:vorticity_dpdy}}
	\caption{\textit{Left}: maximum absolute vorticity at the boundary, rescaled by $\Rey^{-1/2}$, as a function of time $t$ for different Reynolds numbers $\Rey$. The inset shows the same quantity for $\Rey=56~500$, with different curves corresponding to different initial boundary-layer thicknesses. The range of values shown in the inset is highlighted in grey in the main plot.
	\textit{Right}: maximum absolute normal pressure gradient at the boundary, rescaled by $\Rey^{1/2}$, as a function of time $t$ for different Reynolds numbers $\Rey$. The inset shows the peak values as a function of $\Rey$ on logarithmic axes. The black, dashed line indicates the reference scaling~$\Rey^{3/4}$.}
	\label{fig:maximum_vorticity_dp_dy}
\end{figure}

Figure~\ref{fig:maximum_vorticity_dp_dy}~(\textit{Left}) shows the time evolution of the rescaled maximum vorticity in the domain, $\Rey^{-1/2} \sup_\mathcal{D} | \omega|$, for different Reynolds numbers.
We have verified that the maximum vorticity in the domain always occurs at the boundary.
The superposition of the curves at the early times confirms the Prandtl boundary-layer scaling, $\omega|_{\partial \mathcal{D}} \sim \Rey^{1/2}$.
However, this scaling does not persist at later times.
In particular, for $t \approx 1.0$, we observe a deviation from the Prandtl scaling, characterised by a stronger growth of the maximum vorticity.
We further find, although not shown in the figure, that the peak values of the maximum vorticity scale as
\begin{equation}
	\sup_{\partial \mathcal{D}} | \omega| \sim \Rey,
\end{equation}
which is not compatible with the Prandtl boundary-layer expansion.

We remark that the time at which the deviation from the Prandtl scaling occurs, as well as the time at which the maximum vorticity reaches its peak, correlates well with the onset of the first detachments observed in the snapshots.
These numerical observations support the view that the boundary-layer expansion ceases to be valid once the boundary layer separates from the wall, as was previously pointed out in the literature for a different flow configuration~\citep{nguyen2018energy}.

The above conclusions are further supported by measurements of the pressure normal gradient at the boundary, $(\partial p / \partial y )|_{\partial\mathcal{D}}$.
Fig.~\ref{fig:maximum_vorticity_dp_dy}~(\textit{Right}) shows the absolute value of this quantity rescaled by $\Rey^{1/2}$.
Again, the curves for different Reynolds numbers superpose at early times, indicating the expected Prandtl scaling $| \partial p / \partial y | = \mathcal{O}(\Rey^{-1/2})$ at the boundary.
However, this scaling is lost at approximately $t \approx 1$, coinciding with the onset of the first detachments, and the normal pressure gradient instead starts to increase with the Reynolds number.

The inset of Fig.~\ref{fig:maximum_vorticity_dp_dy}~(\textit{Right}) shows the maximum in time of the normal pressure gradient as a function of the Reynolds number.
A best fit to the data reveals the scaling
\begin{equation}\label{EQ:max_dpdy}
	\sup_{\partial \mathcal{D}} | \partial p / \partial y| \sim \Rey^{3/4}.
\end{equation}
This quantity has no \textit{a priori} scaling.
In fact, it is related to the tangential gradient of the vorticity at the boundary.
Considering the wall-normal component of the Navier--Stokes equations and using $\Delta v = \partial \omega / \partial x$, we obtain
\begin{equation}\label{EQ:Lighthill}
	\frac{\partial p}{\partial y} = \frac{1}{\Rey}\frac{\partial \omega}{\partial x} \qquad \text{on} \ \partial \mathcal{D}.
\end{equation}
Equation~\eqref{EQ:Lighthill} is the tangential component of \citet{lighthill1963introduction} vorticity-generation principle, which relates the flux of vorticity through the boundary to the tangential derivative of the pressure.
In view of this relation, the scaling~\eqref{EQ:max_dpdy} provides the strong tangential gradient growth
\begin{equation}
	\sup_{\partial \mathcal{D}} | \partial \omega / \partial x| \sim \Rey^{7/4}.
\end{equation}

Together, the dynamical evolution of both the maximum vorticity and the normal pressure gradient at the boundary provides evidence of the breakdown of the Prandtl boundary-layer expansion at the onset of boundary-layer detachment.

\subsection{Successive detachments and turbulence}

Following the onset of boundary-layer separation, successive detachments occur at different locations along the boundary.
At large Reynolds numbers, this process appears to exhibit a degree of periodicity, which can be seen in the time series of the maximum vorticity and maximum normal pressure gradient at the boundary shown in Fig.~\ref{fig:maximum_vorticity_dp_dy} for the three runs with the highest Reynolds numbers.

The periodic detachments can also be observed in the vorticity snapshots in Fig.~\ref{fig:vorticity_snapshots} for $\Rey=346\,000$ at $t=2.9$, where a stream of small detached vortices is transported from the boundary towards the centre of the domain by the large-scale flow.
At this stage, the bulk flow is strongly affected by the detached vortices, and viscous effects are no longer confined to a thin region near the wall.
This is the ultimate consequence of the breakdown of the Prandtl boundary-layer expansion.
As shown in the full-domain vorticity snapshots in Fig.~\ref{fig:vorticity_snapshots_domain}, the detached vortices reach points at a wide range of distances from the wall and can eventually reach the opposite boundary of the channel, where they interact with it and trigger new boundary-layer detachments.

The advection of the detached vortices by the large-scale flow, together with their mutual interactions, generates a complex turbulent state.
In particular, we observe strong nonlinear effects associated with vortex aggregation and filamentation, which are characteristic mechanisms of two-dimensional hydrodynamic turbulence~\citep{boffetta2012two}.

Turbulent features of the flow can be characterised through the energy spectrum.
However, due to the presence of boundaries, the classical definition of the energy spectrum for homogeneous and isotropic turbulence is not directly applicable.
Instead, we Fourier transform only in the periodic direction, $x$, and obtain $y$-dependent energy spectra. 
To obtain a global characterisation of the flow, we integrate these spectra with respect to $y$ across the full channel and subsequently average them in time over the turbulent regime.
Specifically, we define the energy spectrum as
\begin{equation}
	E(k) = \left\langle \int_{-L/2}^{L/2} \sum_{k \leq k_x < k + 1} | \hat{\bu}(k_x,y)|^2 \, dy \right\rangle.
\end{equation}
where the brackets, $\langle \cdot \rangle$, denote time average over the turbulent regime.

Classical \citet{kraichnan1967inertial}--\citet{leith1968diffusion}--\citet{batchelor1969computation} theory for two-dimensional forced and dissipative turbulence predicts a dual cascade involving the two positive-definite inviscid invariants: an inverse cascade of kinetic energy,
\begin{equation}
	E = \frac{1}{2} \int_\mathcal{D} | \bu(\bx,t)|^2 d \bx,
\end{equation}
and a direct cascade of enstrophy
\begin{equation}
	Z = \frac{1}{2} \int_{\mathcal{D}} | \omega( \bx,t)|^2 d \bx.
\end{equation}
The crossover between the two cascades occurs at the forcing scale $k_f$.
In this scenario, the energy spectrum for the inverse cascade scales as
\begin{equation}
	E \sim \epsilon^{2/3} k^{-5/3} \qquad k_\alpha \ll k \ll k_f,
\end{equation}
where $\epsilon$ is the mean energy dissipation rate, and $k_\alpha$ is the wavenumber associated with the largest-scale damping.
In the direct enstrophy-cascade range, the energy spectrum scales as
\begin{equation}\label{EQ:direct_cascade}
	E \sim \eta^{2/3} k^{-3}, \qquad k_f \ll k \ll k_\eta
\end{equation}
up to logarithmic corrections in the wavenumber~\citep{kraichnan1971inertial}, where $\eta$ is the mean enstrophy dissipation rate, and $k_\eta$ is the Kraichnan dissipation wavenumber given by
\begin{equation}\label{EQ:Kraichnan_scale}
	k_\eta \sim \Rey^{1/2} \eta^{1/6}.
\end{equation}

\begin{figure}
	\centering
	\subfloat{\includegraphics[width=0.85\columnwidth]{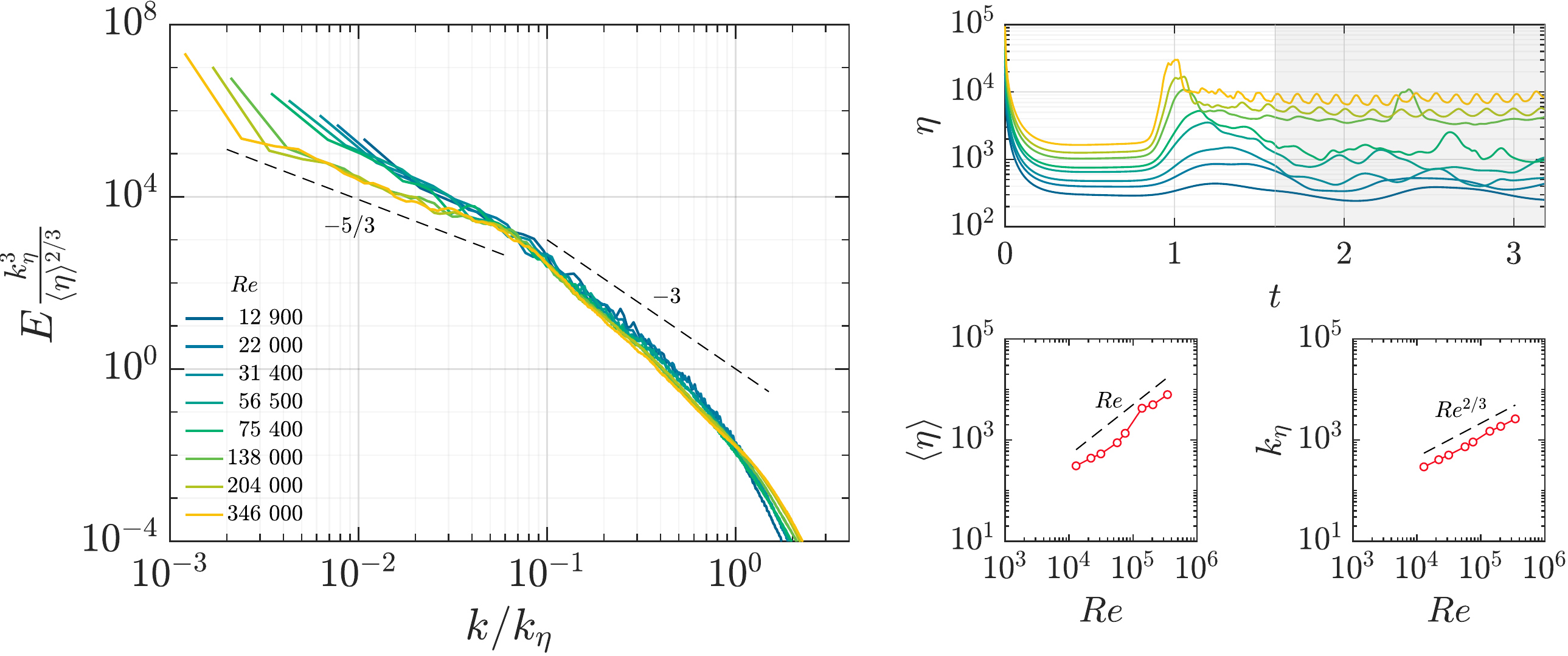}\label{fig:spectra}}
	\caption{\textit{Left:}~Time-averaged energy spectra~$E$, compensated by~$k_\eta^3/ \langle \eta \rangle^{2/3}$, as a function of the rescaled wavenumber~$k / k_\eta$ for different Reynolds numbers~$\Rey$.
	\textit{Right,~top:}~Time evolution of the enstrophy dissipation rate~$\eta$ for different Reynolds numbers.
	\textit{Right,~bottom:}~Time-averaged enstrophy dissipation rate~$\langle \eta \rangle$ and Kraichnan dissipation wavenumber~$k_\eta$ as functions of the Reynolds number~$\Rey$. The time averages are computed over the turbulent regime, within the time interval highlighted in grey in the top-right panel.}
\end{figure}

This dual-cascade scenario has also been investigated in wall-bounded flows, where the inverse cascade appears to be particularly pronounced near the boundaries~\citep{clercx2000energy,clercx2005no,kramer2008large,hazi2010lattice}.
In our flow configuration, however, the enstrophy dissipation rate has a strong dependence on the Reynolds number due to enstrophy generation at the boundaries.
Fig.~\ref{fig:spectra}~(\textit{Right,~top}) shows the instantaneous enstrophy dissipation rate,
\begin{equation}
	\eta(t) = \frac{1}{\Rey} \int_\mathcal{D} | \nabla \omega (\bx,t)|^2 d \bx.
\end{equation}
Similarly to the other quantities discussed above, $\eta(t)$ reaches a peak at the onset of boundary-layer detachment, followed by a complex, turbulent evolution.
We observe that the mean enstrophy dissipation rate in the turbulent regime (highlighted in grey) increases with the Reynolds number.
Indeed, Fig.~\ref{fig:spectra}~(\textit{Right,~bottom}) shows the time average $\langle\eta\rangle$ in the turbulent regime as a function of the Reynolds number.
The data are consistent with the scaling
\begin{equation}\label{EQ:eta_scaling}
	\langle \eta \rangle \sim \Rey.
\end{equation}
Using this scaling in Eq.~\eqref{EQ:Kraichnan_scale}, we obtain the Reynolds-number dependence of the Kraichnan dissipation wavenumber,
\begin{equation}\label{EQ:k_eta_scaling}
	k_\eta \sim \Rey^{2/3}.
\end{equation}
This scaling is verified in Fig.~\ref{fig:spectra} (\textit{Right,~bottom}).

Figure~\ref{fig:spectra} (\textit{Right}) shows the compensated energy spectrum $E k_\eta^3 / \langle \eta \rangle ^{2/3}$.
This compensation is motivated by the direct enstrophy-cascade scaling~\eqref{EQ:direct_cascade}, which provides
\begin{equation}
	\frac{Ek_\eta^3}{\langle \eta \rangle ^{2/3}} \sim \left(\frac{k}{k_\eta} \right)^{-3}
\end{equation}
in the enstrophy direct-cascade range.
In view of scalings~\eqref{EQ:eta_scaling} and~\eqref{EQ:k_eta_scaling}, the energy spectrum is compensated by $k_\eta^3 / \langle \eta \rangle^{2/3} \sim \Rey^{4/3}$.
The compensated energy spectra for different Reynolds numbers collapse well over a broad range of scales.
In particular, the spectra exhibit two scaling ranges consistent with a dual-cascade interpretation: below a crossover wavenumber $k_f$, the energy spectrum scales approximately as $E\sim k^{-5/3}$, indicating an inverse energy cascade, while above $k_f$ it follows a steeper power law, $E\sim k^{-\gamma}$, with $\gamma$ slightly larger than $3$, indicating a direct enstrophy cascade.
Previous numerical studies of two-dimensional turbulence have likewise reported a slightly steeper power law in the direct-cascade range, with the exact $k^{-3}$ scaling being recovered only in the high-resolution, large-Reynolds-number limit~\citep{boffetta2010evidence}.

We also note that, although the theory of dual cascades in two-dimensional turbulence is traditionally formulated for statistically stationary, forced and dissipative flows, a similar cascade scenario has also been observed numerically in freely decaying two-dimensional turbulence~\citep{mininni2013inverse}.
In flows bounded by solid walls, the crossover scale between the inverse energy cascade and the direct enstrophy cascade has been related to the mean boundary-layer thickness~\citep{clercx2000energy}.

The energy spectrum, however, is not homogeneous along the wall-normal direction.
From the snapshots of Fig.~\ref{fig:vorticity_snapshots}, we observe that the detached vortices are not uniformly distributed on the domain, with regions close to the boundary exhibiting stronger nonlinear interactions.

\begin{figure}
	\centering
	\subfloat{\includegraphics[width=0.9\columnwidth]{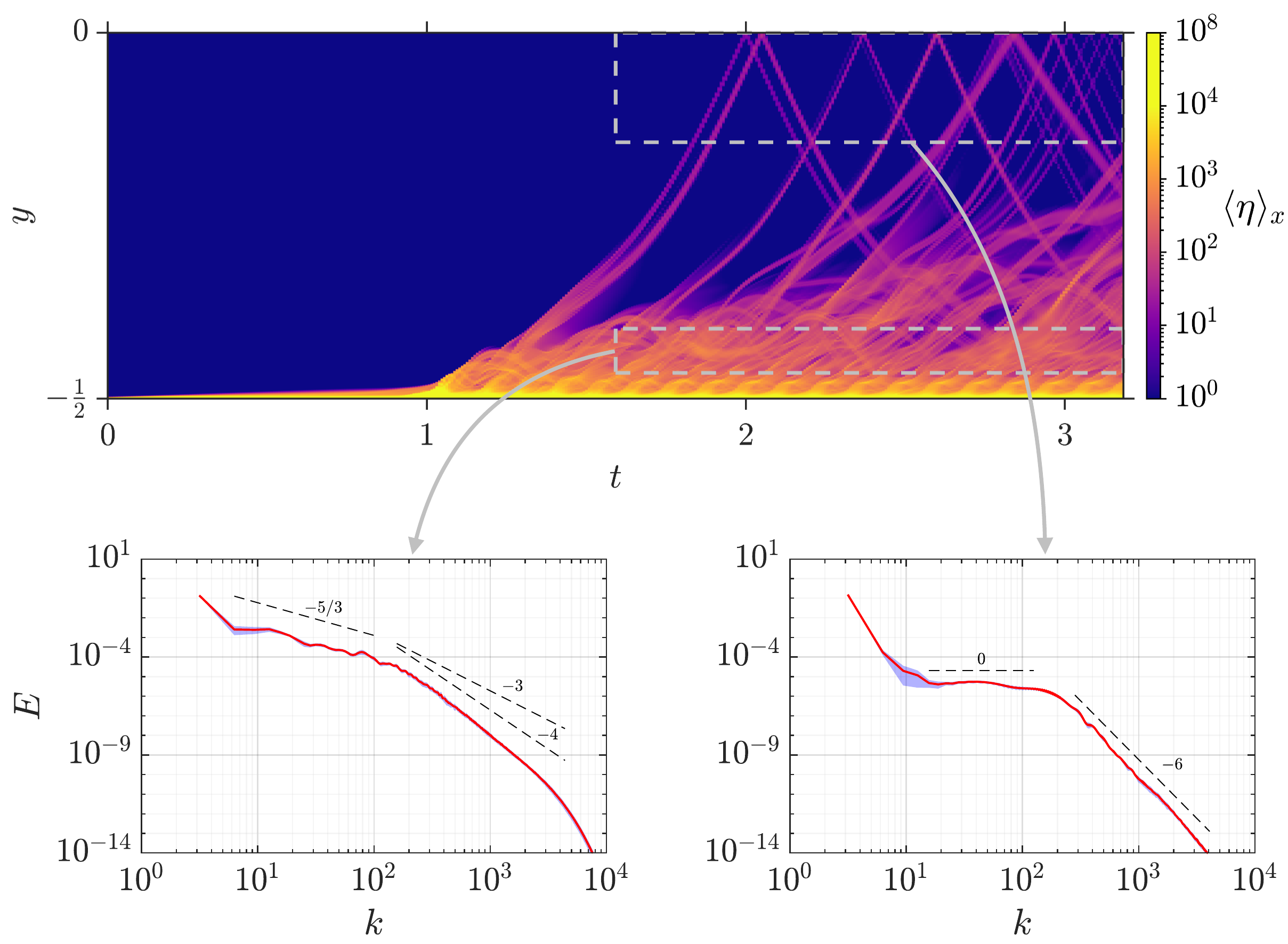}}
	\caption{\textit{Top:} Colormap of $x$-averaged enstrophy dissipation rate, $\langle \eta \rangle_x$, as a function of time $t$ and the wall-normal coordinate $y$. Owing to the symmetry of the flow, only half of the $y$-domain is shown. \textit{Bottom, left:} Energy spectrum averaged over time and across the wall-normal range near the boundary during the turbulent regime. \textit{Bottom, right:} Energy spectrum averaged over time and across the wall-normal range away from the boundary during the turbulent regime. In both spectra, the red line indicates the mean spectrum, while the blue shaded area represents the standard deviation across the corresponding wall-normal range. The black dashed lines indicate the reference scalings. The wall-normal ranges and time interval used to compute the spectra are indicated by grey dashed lines in the top panel.}
	\label{fig:regions}
\end{figure}

These features are better characterised through the $x$-averaged enstrophy dissipation rate, $\langle \eta \rangle_x$ as a function of $y$ and $t$, shown in Fig.~\ref{fig:regions}~(\textit{Top}).
At early times, strong enstrophy dissipation is confined to a narrow region near the wall, corresponding to the attached boundary layer.
Following the onset of boundary-layer detachments at $t \approx 1$, we observe a complex redistribution of enstrophy dissipation from the boundary into the bulk, which separates the domain into distinct regions.
Close to the boundary, the flow exhibits strong turbulent activity, characterised by an approximately locally homogeneous enstrophy dissipation rate.
Away from the boundary, the flow is dominated by the ballistic motion of individual, weakly correlated vortices.
These two regions are delineated by the grey boxes in the figure.

Accordingly, the two regions exhibit markedly different energy spectra.
Fig.~\ref{fig:regions}~(\textit{Bottom,~left}) shows the energy spectrum averaged both in time and over $y$ in the region close to the boundary.
A clear dual-cascade scenario is observed:
at large scales, the spectrum exhibits a $k^{-5/3}$ slope, consistent with an inverse cascade of energy, while at smaller scales it follows a steeper slope, close to $k^{-4}$, consistent with the direct cascade of enstrophy in a finite-resolution simulation.

Figure~\ref{fig:regions}~(\textit{Bottom,~right}) shows the energy spectrum averaged both in time and over $y$ in the region away from the boundary.
Here, the uncorrelated large-scale vortices produce a characteristic $k^0$ spectrum, while the smaller detached vortices exhibit greater regularity, with a steeper spectrum scaling as $k^{-6}$. A spectrum $k^{-\alpha}$ with $\alpha < 3$ is typical of a direct enstrophy cascade contaminated by the large-scale vortices~\citep{benzi1986intermittency,reynoso2024self}.

\subsection{How the boundary-layer detachments depend on the initial boundary-layer thickness $\delta$}

One may ask how all the quantities measured above depend on the initial boundary-layer thickness or, more generally, on the choice of the initial boundary-layer profile.
The inset of Fig.~\ref{fig:maximum_vorticity_dp_dy}~(\textit{Left}) shows the time evolution of the maximum vorticity, rescaled by $\Rey^{-1/2}$, for different initial boundary-layer thicknesses.
Different curves correspond to distinct initial boundary-layer thicknesses, while the Reynolds number is kept fixed at $\Rey = 56\,500$.
We observe that all curves rapidly converge to the same value, with thinner initial boundary layers converging more rapidly.
We have also observed, although not shown, that the boundary-layer profiles converge on comparable time scales.
These observations suggest that the influence of the initial boundary-layer thickness and, more generally, of the initial boundary-layer profile on the subsequent evolution is weak.
We therefore conjecture that the qualitative features described above are largely insensitive to the choice of the initial boundary-layer conditions.

All the phenomenological observations presented above highlight the complexity of the flow generated by the detached vortices.
We now turn to the infinite-Reynolds-number limit and investigate it with the aid of rigorous mathematical criteria.

\section{The mathematics of convergence}
\label{sec:maths}

Given the complexity of the detached vortices described in the previous section, one may ask \textit{whether the Navier--Stokes solutions converge to the simple stationary Taylor--Green solution of the Euler equations in the infinite-Reynolds-number limit}.
We begin by formulating this question in terms of a precise topology for the convergence.

In what follows, we understand the infinite-Reynolds-number limit in the energy space, uniformly on the compact time interval $[0,T]$. More precisely, we say that the Navier--Stokes solutions converge to the Euler solution if
\begin{equation}
	E_\Delta(t) = \frac{1}{2}\int_{\mathcal{D}} | \bu(\bx,t) - \overline{\bu}(\bx)|^2 \, d\bx \to 0 \qquad \text{uniformly in } t \in [0,T],
\end{equation}
as $\Rey \to \infty$.
The quantity $E_\Delta$ measures the energy of the difference between the two solutions.
We fix $T$ as the time horizon of our simulations, chosen to be approximately three times the time required for the first strong detachments to occur.
At $t\approx T$, the detached vortices generated at one boundary have already been transported by the large-scale flow to the opposite boundary, with which they have started to interact.

\subsection{Direct computation of convergence norm}

The first measurements concern the direct numerical computation of the convergence norm.
Fig.~\ref{fig:convergence_norm_dissipation}~(\textit{Left}) shows the time evolution of $E_\Delta$, rescaled by $\Rey^{1/2}$, for different Reynolds numbers.
At early times, the curves superpose almost perfectly, with the energy $E_\Delta$ growing exponentially in time, but monotonically decreasing as $\Rey^{-1/2}$ as the Reynolds number increases.
The rate of exponential growth in time is independent of the Reynolds number.
In this regime, the difference between the Navier--Stokes and Euler solutions is dominated by the energy contained in the attached boundary layer, for which Prandtl boundary-layer theory predicts convergence to zero at the rate $\Rey^{-1/2}$.

At later times, the growth of $E_\Delta$ saturates and oscillations emerge.
These oscillations arise from the complex turbulent dynamics generated by the detached vortices.
Despite this more complex behaviour, $E_\Delta$ still appears to scale as $\Rey^{-1/2}$ throughout the oscillations.
This observation provides numerical evidence that the Navier--Stokes solutions continue to converge to the strong Euler solution, even after the onset of boundary-layer detachment.

Further insights can be obtained from \textit{a priori} estimates for the quantity $E_\Delta$.
Using the Navier--Stokes and Euler equations, we obtain the following inequality:
\begin{equation}\label{EQ:a_priori_estimate_first}
	\frac{dE_\Delta}{dt} \leq 2 E_\Delta \, \sup_{\mathcal{D}} | \nabla \overline{\bu}| + \frac{1}{\Rey} \int_{\mathcal{D}} (\bu - \overline{\bu}) \cdot \Delta \bu \, d\bx.
\end{equation}
The first term on the right-hand side arises from the nonlinearities, while the second accounts for the viscous contribution in the Navier--Stokes equations.
The pressure terms do not contribute, because of the incompressibility of both flows.
We note that the above inequality is well defined because the solutions are smooth; in particular, the Euler solution has bounded gradients.

Applying Grönwall's inequality, the nonlinear contribution gives rise to an exponential upper bound, consistent with the exponential growth of $E_\Delta$ observed in Fig.~\ref{fig:convergence_norm_dissipation}~(\textit{Left}).
The dependence on $\Rey^{-1/2}$ is inherited from the initial discrepancy between the Navier--Stokes and Euler solutions and is consistent with the convergence rate observed numerically.

The last term in Eq.~\eqref{EQ:a_priori_estimate_first} plays a crucial role in determining whether convergence occurs, as it cannot be bounded \textit{a priori} in a way that is uniform in the Reynolds number.
Several approaches to establishing convergence criteria have been proposed, many of which focus on how to control this term.
In what follows, we consider some of these criteria and apply them to the present flow, with particular emphasis on the complex and non-intuitive turbulent regime.

\subsection{Kato criterion for energy dissipation rate}

\begin{figure}
	\centering
	\subfloat{\includegraphics[width=0.46\columnwidth]{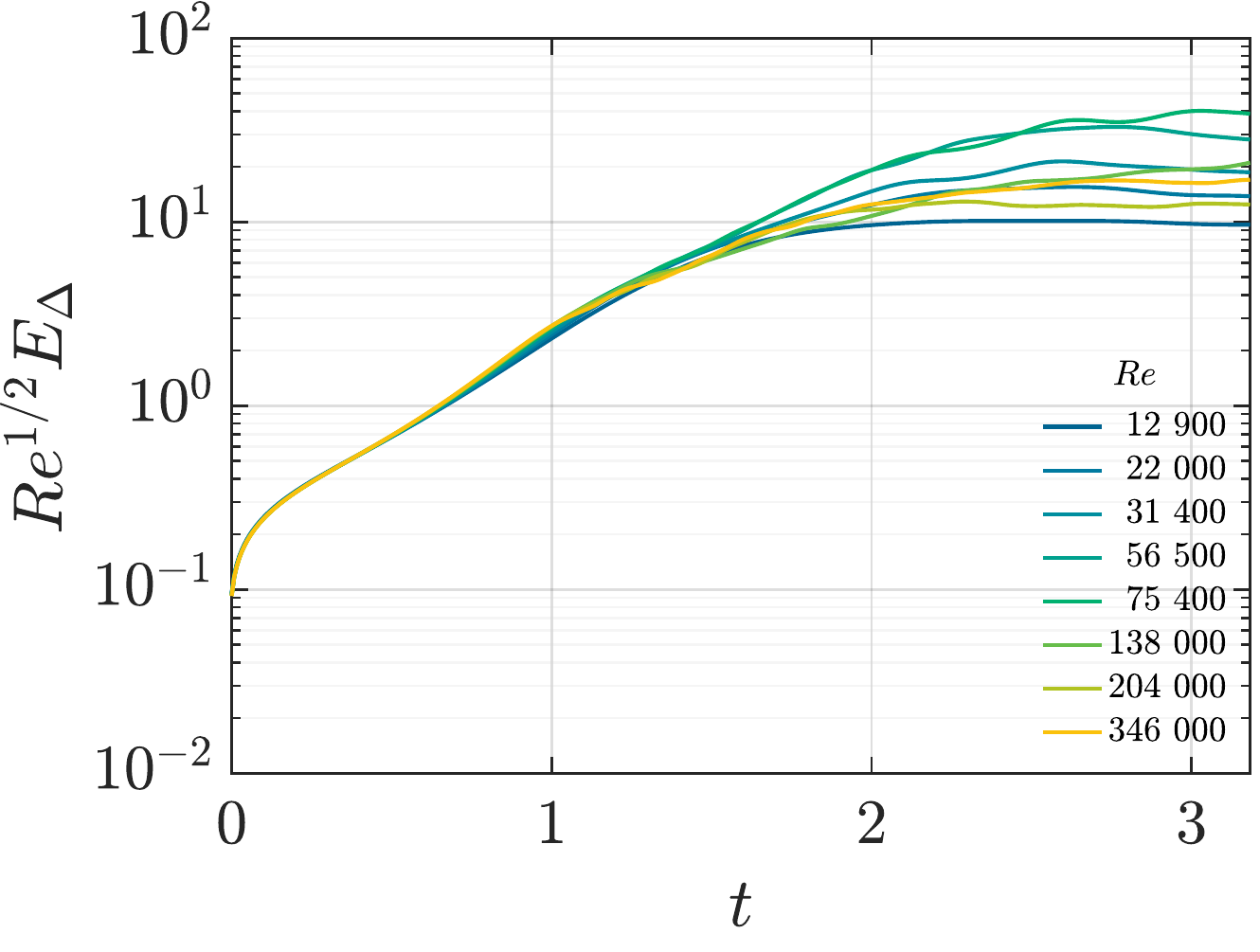}\label{fig:L2_convergence}}
	\qquad
	\subfloat{\includegraphics[width=0.46\columnwidth]{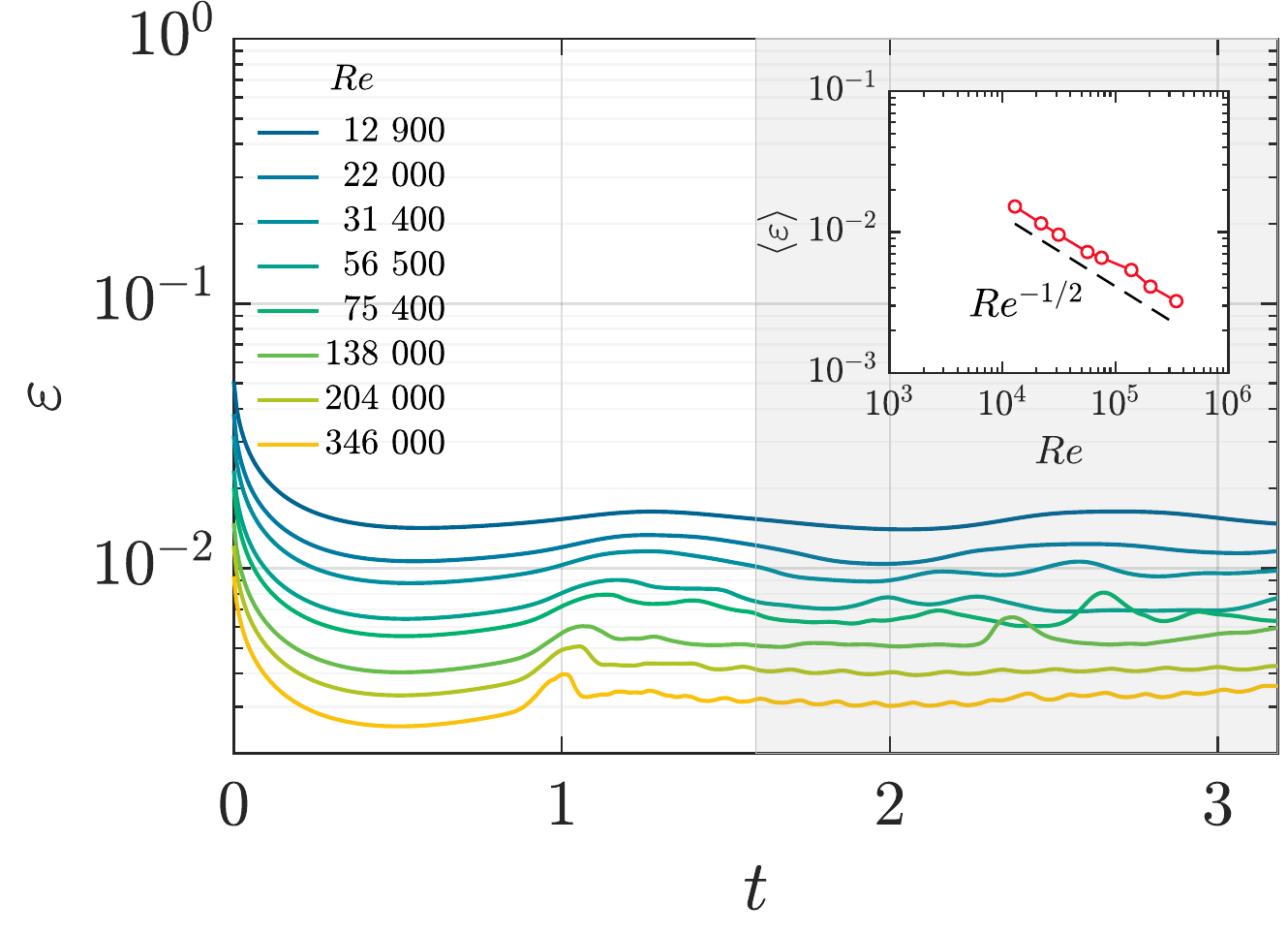}\label{fig:dissipation_criterion}}
	\caption{\textit{Left:}~Energy of the difference between Navier--Stokes and Euler solutions,~$E_\Delta$, rescaled by $\Rey^{1/2}$, as a function of time $t$ for different Reynolds numbers~$\Rey$.
	\textit{Right:}~Time evolution of the total energy dissipation rate $\varepsilon$ for different Reynolds numbers. The inset shows the time-averaged energy dissipation rate $\langle \varepsilon \rangle$ as a function of the Reynolds number in logarithmic axes. The black dashed line indicates the reference scaling $\Rey^{-1/2}$. The time average is computed over the turbulent regime, highlighted in grey in the main plot.}
	\label{fig:convergence_norm_dissipation}
\end{figure}

The most classical convergence criterion is due to~\cite{kato1984remarks}, who proved that the convergence takes place if and only if the mean energy dissipation rate vanishes in the infinite-Reynolds-number limit.
Moreover, he showed that it is sufficient to consider the energy dissipation in a layer close to the boundary whose thickness scales as $\Rey^{-1}$ rather than in the entire domain.
In symbols, the convergence takes place if and only if
\begin{equation}\label{EQ:Kato_criterion}
	\left\langle \frac{1}{\Rey}\int_{\Gamma_{\Rey^{-1}}} | \nabla \bu|^2 \, d\bx \right\rangle \to 0, \qquad \text{as} \ \Rey \to \infty,
\end{equation}
where
\begin{equation}
	\Gamma_{\Rey^{-1}} = \left\{ \bx \in \mathcal{D} : \mathrm{dist}(\bx,\partial \mathcal{D}) \leq c \Rey^{-1} \right\},
\end{equation}
and $c>0$ is an arbitrary fixed constant.

Necessity of condition~\eqref{EQ:Kato_criterion} follows from the energy inequality of Leray--Hopf solutions.
To prove sufficiency, Kato introduced suitable boundary-layer correctors to the Euler solution.
The last term of the inequality~\eqref{EQ:a_priori_estimate_first} can then be estimated in terms of the gradients $\nabla \bu$, which are not bounded \textit{a priori}.
In the argument, the only boundary-layer thickness of his correctors which are compatible with condition~\eqref{EQ:Kato_criterion} is of order $\mathcal{O}(\Rey^{-1})$.

Kato's criterion has led to a substantial body of work, including improvements to the original criterion~\citep{wang2001kato} and reformulations in terms of other quantities that must vanish in the infinite-Reynolds-number limit~\citep{temam1997behavior,kelliher2007kato}.

From a physical perspective, Kato's criterion links convergence to the absence of anomalous energy dissipation in the infinite-Reynolds-number limit.
In particular, if anomalous dissipation occurs, it must be associated with strong dissipation events within a layer of thickness $\mathcal{O}(\Rey^{-1})$, which is thinner than the Prandtl boundary layer.
It is worth emphasising, however, that the $\mathcal{O}(\Rey^{-1})$ Kato layer is primarily a mathematical construct~\citep{lopes2019fluids}, and its precise physical significance remains unclear.

We begin by studying the total energy dissipation rate
\begin{equation}\label{EQ:total_dissipation}
	\varepsilon(t) = \frac{1}{\Rey}\int_{\mathcal{D}} | \nabla \bu (\bx,t)|^2 \, d\bx,
\end{equation}
whose time evolution is shown in Fig.~\ref{fig:convergence_norm_dissipation}~(\textit{Right}).
At early times, we observe a monotonic decrease of $\varepsilon$, followed by pronounced peaks at the onset of boundary-layer separation and a subsequent turbulent, oscillatory regime.
At later times, monotonicity breaks down, with strong dissipation events emerging at certain Reynolds numbers.
Nevertheless, the time-averaged dissipation rate in the turbulent regime, shown in the inset, decreases as $\Rey^{-1/2}$ as $\Rey\to\infty$.
Remarkably, the Prandtl scaling persists even after the onset of boundary-layer detachment.
We have also verified the same scaling both before and during the dissipation peaks.

To study the behaviour of the dissipation rate close to the boundary, we introduce the wall-normal distance $y^\ast$ and consider the layer-averaged energy dissipation rate
\begin{equation}
	\langle \varepsilon \rangle_{\Gamma_{y^\ast}} = \left\langle \frac{1}{\Rey}\int_0^{y^\ast}\int \left| \nabla \bu \left(x,y-\frac{1}{2},t \right) \right|^2 \, dx \, dy \right\rangle,
\end{equation}
which is the time-averaged energy dissipation rate within a layer of thickness $y^\ast$ adjacent to the bottom boundary.

\begin{figure}
	\centering
	\subfloat{\includegraphics[width=0.46\columnwidth]{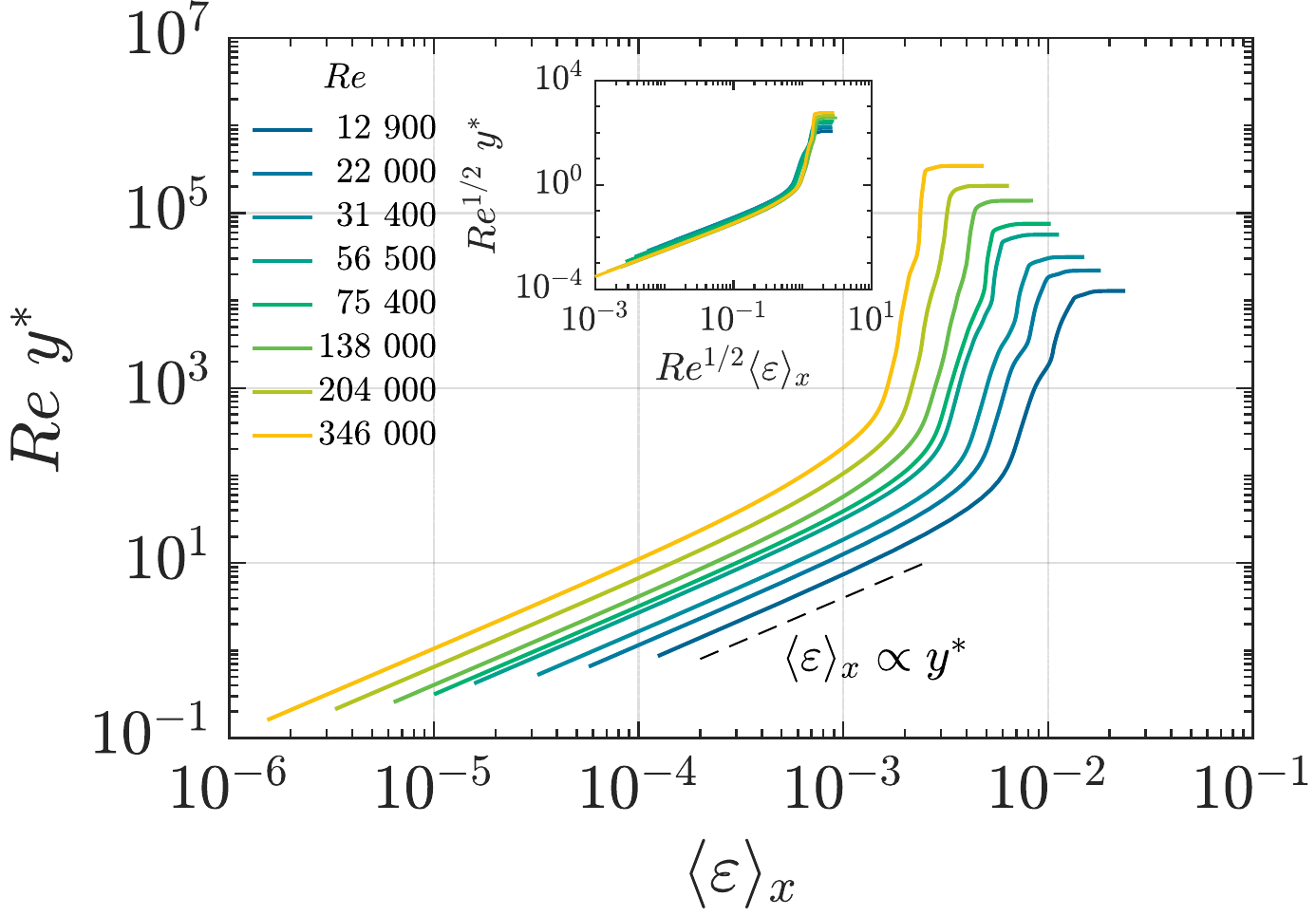}\label{fig:kato_criterion}}
	\qquad
	\subfloat{\includegraphics[width=0.46\columnwidth]{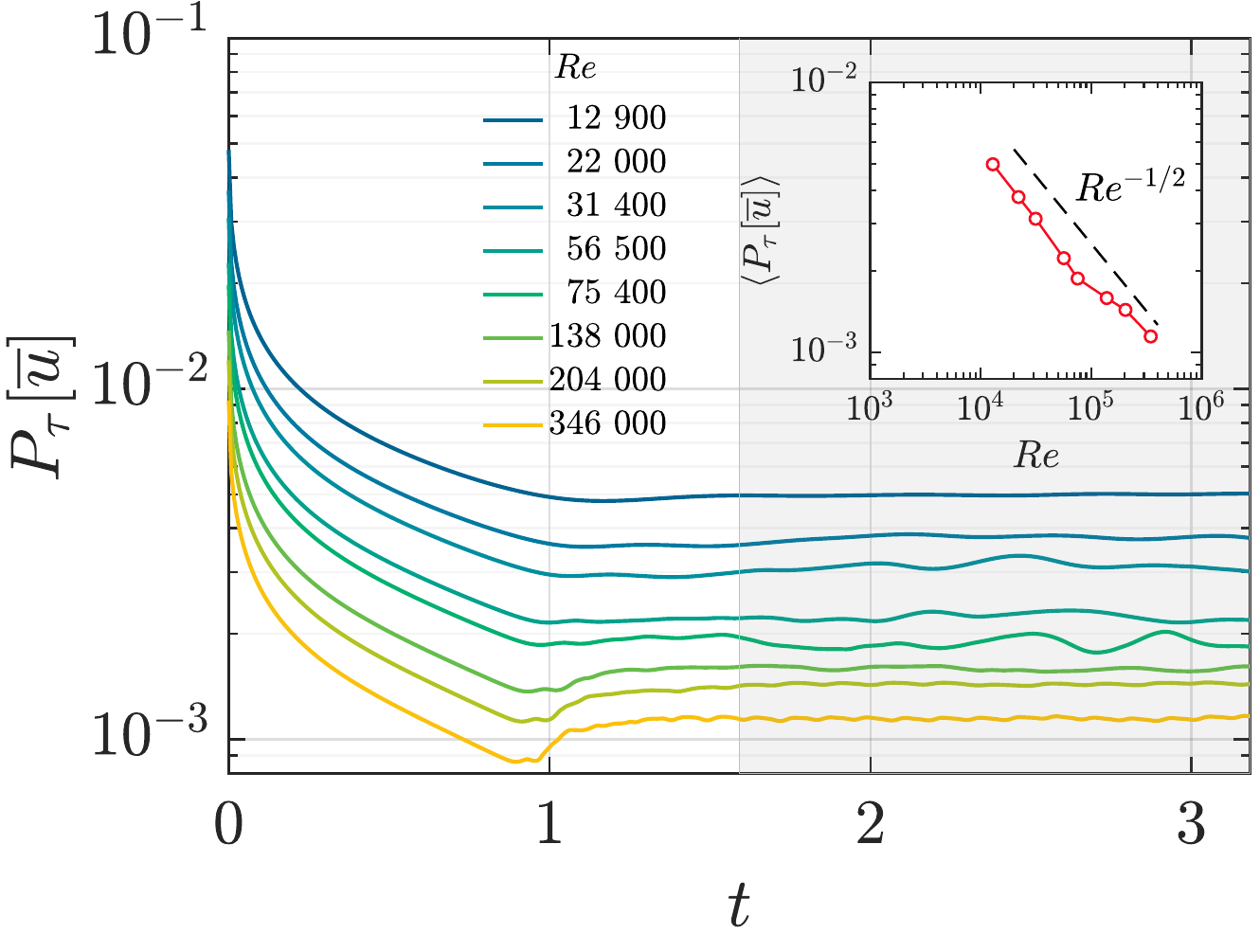}\label{fig:bardos_criterion}}
	\caption{\textit{Left:}~Wall-normal distance $y^\ast$, rescaled by $\Rey$, as a function of the energy dissipation rate averaged over the strip $\Gamma_{y^\ast}$, $\langle \varepsilon \rangle_{\Gamma_{y^\ast}}$, for different Reynolds numbers $\Rey$. The black dashed line indicates the reference scaling $\langle \varepsilon \rangle_{\Gamma_{y^\ast}} \propto y^\ast$ in the viscous sublayer. The inset shows $y^\ast$ as a function of $\langle \varepsilon \rangle_{\Gamma_{y^\ast}}$, with both rescaled by $\Rey^{1/2}$.
	\textit{Right:}~Time evolution of the power associated with viscous shear stresses evaluated against the Euler slip velocity, $P_\tau[\overline{\bu}]$, for different Reynolds numbers $\Rey$. The inset shows the time-averaged power, $\langle P_\tau [\overline{\bu}] \rangle$, as a function of the Reynolds number on logarithmic axes. The black dashed line indicates the reference scaling $\Rey^{-1/2}$. The time average is computed over the turbulent regime, highlighted in grey in the main plot.}
	\label{fig:Kato_layer_Kelliher}
\end{figure}

Figure~\ref{fig:Kato_layer_Kelliher}~(\textit{Left}) shows the wall-distance $y^\ast$, rescaled by $\Rey$, as a function of $\langle \varepsilon \rangle_{\Gamma_{y^\ast}}$ for different Reynolds numbers.
At small values of $y^\ast$, all curves exhibit the linear scaling law
\begin{equation}
	\langle \varepsilon \rangle_{\Gamma_{y^\ast}} \propto y^\ast.
\end{equation}
This is the expected behaviour within the viscous sublayer, where the velocity field depends linearly on the wall-normal distance~\citep{landau1959fluid}.
At larger values of $y^\ast$, the viscous sublayer gives way to the turbulent layer populated by detached vortices.

The layer-averaged dissipation rate decreases monotonically with increasing Reynolds number for all rescaled wall-normal distances.
The inset shows $y^\ast$, rescaled by $\Rey^{1/2}$, as a function of $\langle\varepsilon\rangle_{\Gamma_{y^\ast}}$, also rescaled by $\Rey^{1/2}$.
The collapse of the curves for different Reynolds numbers indicates that the dissipation follows the $\Rey^{-1/2}$ scaling associated with the Prandtl boundary layer.
In particular, at a wall-normal distance of order $\Rey^{-1}$, the dissipation vanishes in the infinite-Reynolds-number limit, indicating that the Kato criterion for convergence is satisfied.

\subsection{Bardos--Titi--Kelliher criterion for the power of viscous shear stresses}

While Kato's condition involves measurements in a thin layer near the boundary, there is also a convergence criterion formulated directly in terms of quantities evaluated at the boundary itself.
Integrating by parts the last term in inequality~\eqref{EQ:a_priori_estimate_first} yields a boundary contribution that controls convergence.
More precisely, the Navier--Stokes solutions converge to the Euler solution if and only if
\begin{equation}
	\left\langle \frac{1}{\Rey} \int_{\partial \mathcal{D}} \overline{\bu}_\tau \, \omega \, dS \right\rangle_t \to 0, \qquad \text{uniformly in } t \in [0,T], \quad \text{as} \ \Rey \to \infty,
\end{equation}
where the brackets $\langle\cdot\rangle_t$ denote a time average over the interval $[0,t]$, and $\overline{\bu}_\tau$ denotes the tangential component of the Euler solution at the boundary.
In our flow configuration, this coincides with the $x$-component $\overline{u}$ on the bottom wall (and $-\overline{u}$ on the top wall). Its trace is given explicitly in~\eqref{EQ:Euler_slip}, in dimensional form.

This statement was, to the best of our knowledge, first proved in an equivalent form by~\cite{bardos2013mathematics}; see also~\citep{bardos2018remarks}.
The formulation used here was subsequently presented in~\citep{kelliher2017observations}.

The quantity
\begin{equation}\label{EQ:power_stresses}
	P_\tau [\overline{\bu}] = \frac{1}{\Rey} \int_{\partial \mathcal{D}} \overline{\bu}_\tau \omega \, dS
\end{equation}
appearing in the time average admits a natural physical interpretation as the power associated with the viscous shear stresses evaluated against the Euler slip velocity.
Indeed, the shear stress at the boundary is, in dimensionless form, given by $-\frac{1}{\Rey}\partial u/\partial y$, which, at the boundary, coincides with $\frac{1}{\Rey}\omega$.
Thus, the Bardos--Titi--Kelliher criterion states that the Navier--Stokes solutions converge to the strong Euler solution in the infinite-Reynolds-number limit if and only if the power associated with the viscous stresses acting against the Euler slip flow vanishes uniformly in time.

Since we have the explicit expression~\eqref{EQ:Euler_slip} for the stationary trace of the Euler solution, we can further simplify~\eqref{EQ:power_stresses}.
Using Parseval's identity and the symmetry between the top and bottom boundaries, we obtain, in dimensionless variables,
\begin{equation}
	P_\tau[\overline{\bu}] = -\frac{2}{\Rey} \int_0^2 \cos (\pi x) \omega(x,y=-1/2) \, dx = - \frac{4}{\Rey}  \mathcal{R} [\hat{\omega}(k = 1)],
\end{equation}
where $\mathcal{R}(\cdot)$ denotes the real part and $\hat{\omega}$ is the one-dimensional Fourier transform of the vorticity trace at the bottom boundary, $y=-1/2$.
Therefore, convergence is controlled by a single large-scale Fourier mode of the boundary vorticity, namely $\hat{\omega}(k=1)$.
In particular, failure of convergence requires this mode to produce a non-vanishing contribution to the time-averaged boundary power.
Since $P_\tau[\overline{\bu}]$ is proportional to $\Rey^{-1}\hat{\omega}(k=1)$, this requires the relevant contribution of the $k=1$ boundary-vorticity mode to grow at least as fast as $\Rey$ in the infinite-Reynolds-number limit.
This illustrates the strength of the criterion and the advantage of having an explicit expression for the stationary Euler solution for its interpretation.

Figure~\ref{fig:Kato_layer_Kelliher}~(\textit{Right}) shows the time evolution of~$P_\tau[\overline{\bu}]$ for different Reynolds numbers.
We first observe that this quantity remains positive throughout the entire time interval.
Therefore, there are no cancellations in the time average, and we can characterise its behaviour using the average over the full simulation interval.
We further observe that $P_\tau[\overline{\bu}]$ decreases monotonically in time.
The inset shows its time-averaged value over the turbulent interval highlighted in grey as a function of the Reynolds number.
The data are consistent with a $\Rey^{-1/2}$ decay, which satisfies the Bardos--Titi--Kelliher criterion for convergence.

This scaling is also consistent with the direct measurements of the convergence norm, which decays as $\Rey^{-1/2}$ and is bounded by the corresponding time-averaged boundary power, $\langle P_\tau[\overline{\bu}]\rangle_t$.
We have also computed time averages over the intervals surrounding the dissipation peaks and find the same $\Rey^{-1/2}$ scaling.

% \subsection{Drivas--Nguyen criterion for wall-normal velocity}

% \clearpage

\section{Interpreting observations in terms of vorticity distribution}
\label{sec:vorticity}

The measurements presented in~\S\ref{sec:maths} provide numerical evidence for the convergence of the Navier--Stokes solutions to the simple strong and stationary Euler solution.
Vorticity snapshots, on the other hand, reveal detached vortices that penetrate into the bulk, with increasingly intense and numerous detachments as the Reynolds number increases.
The main goal of this section is to provide an interpretation of these phenomenological observations and to explain how they are consistent with convergence in the high-Reynolds-number limit.
To this end, we exploit detailed distributions of the vorticity and related quantities.

\subsection{Distribution of local energy dissipation}

The detached vortices in the snapshots of Fig.~\ref{fig:vorticity_snapshots} become more numerous and more intense as the Reynolds number increases.
Nevertheless, their typical size also decreases, and the region of the bulk affected by them appears to become increasingly concentrated.
These observations suggest that the total area occupied by the detached vortices decreases with the Reynolds number, which may contribute to the vanishing of the dissipation rate in the high-Reynolds-number limit.

A way to quantify this observation is through the probability density functions (PDFs) of the local dissipation,
\begin{equation}\label{EQ:local_dissipation}
	\varepsilon_{\mathrm{loc}}(\bx, t) = \frac{2}{\Rey}S_{ij}S_{ij},
\end{equation}
where repeated indices are summed over, and
\begin{equation}
	S_{ij} = \frac{1}{2} \left( \frac{\partial u_i}{\partial x_j} + \frac{\partial u_j}{\partial x_i} \right)
\end{equation}
is the rate-of-strain tensor.

We recall two important properties of the local dissipation~\eqref{EQ:local_dissipation}.
First, for an incompressible flow, the difference between $\varepsilon_\mathrm{loc}$ and $\Rey^{-1} | \nabla \bu (\bx,t)|^2$ can be written in divergence form
\begin{equation}
	\varepsilon_{\mathrm{loc}}(\bx,t) = \frac{1}{\Rey} | \nabla \bu |^2 + \frac{1}{\Rey} \nabla \cdot \left( \bu \cdot \nabla \bu \right).
\end{equation}
Under the boundary conditions considered here, the divergence term integrates to zero over the spatial domain.
Consequently, the integral of the local dissipation coincides with the total dissipation~\eqref{EQ:total_dissipation},
\begin{equation}
	\int_{\mathcal{D}} \varepsilon_{\mathrm{loc}}(\bx,t) \, d\bx = \varepsilon(t).
\end{equation}
Second, because it is defined solely in terms of the rate-of-strain tensor, the local dissipation does not include spurious contributions from rigid-body rotations.
For this reason, $\varepsilon_{\mathrm{loc}}$ is also referred to as the \textit{true dissipation}, while $\Rey^{-1}| \nabla \bu(\bx,t)|^2$ is called \textit{pseudo-dissipation}~\citep[\S5.3]{pope2000turbulent}.

\begin{figure}
	\centering
	\includegraphics[width=1\columnwidth]{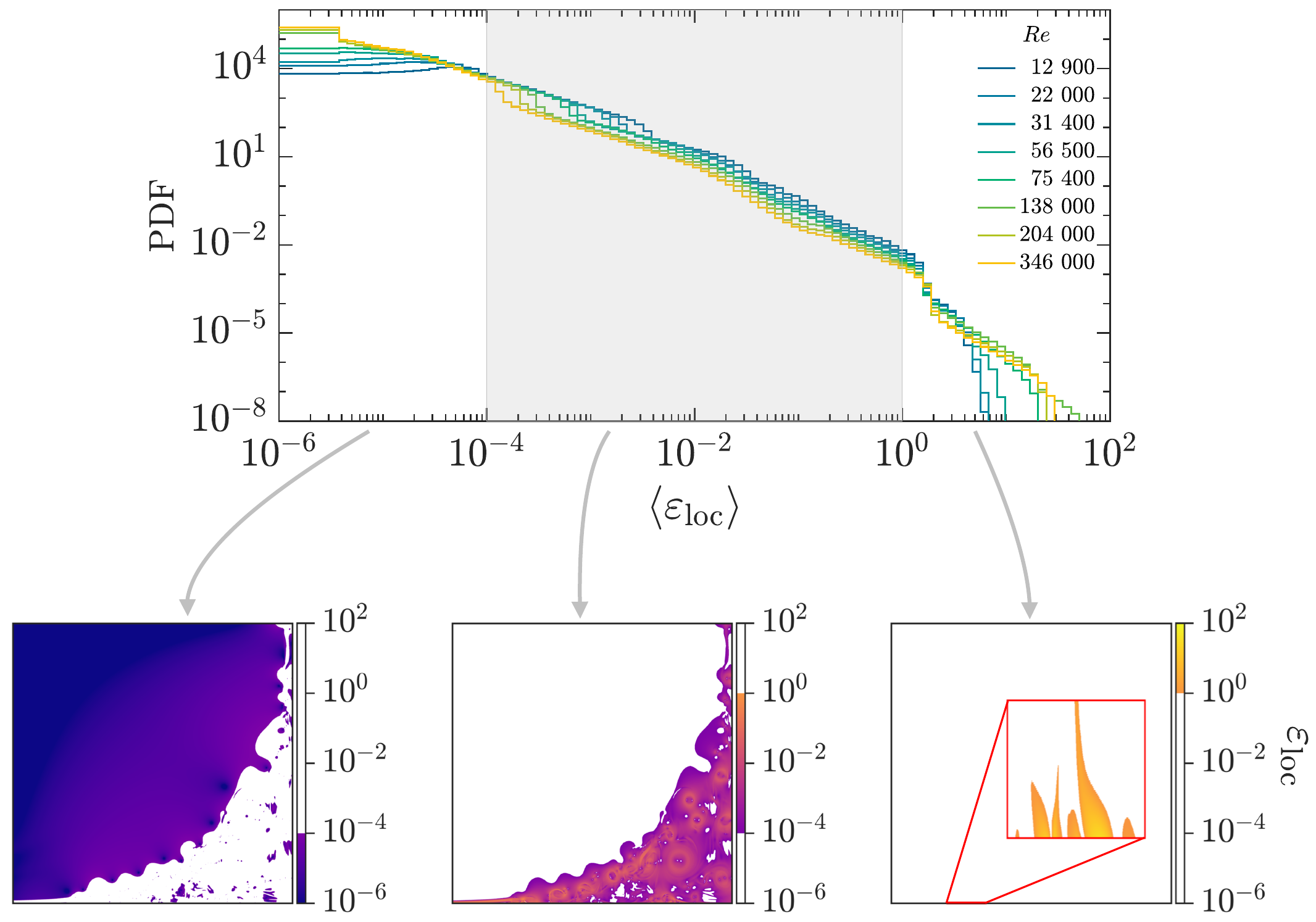}
	\caption{\textit{Top:} Probability density functions (PDFs) of time-averaged local energy dissipation rate, $\langle \varepsilon_\mathrm{loc} \rangle$, for different Reynolds numbers. \textit{Bottom:} Snapshots of local energy dissipation rate $\varepsilon_{\mathrm{loc}}$ for Run VIII, with $\Rey = 346\,000$, at $t = 2.9$. Each snapshot corresponds to a range of local energy dissipation rates indicated in the top panel. In the final snapshot, the red box highlights a close-up of a near-boundary region measuring $7 \times 10^{-2}L$ in the streamwise direction and $9 \times 10^{-4} L$ in the wall-normal direction.}
	\label{fig:dissipation_measures}
\end{figure}

Figure~\ref{fig:dissipation_measures} shows the PDFs of $\varepsilon_{\mathrm{loc}}$, averaged in time over the turbulent regime, for different Reynolds numbers.
From these distributions, we can distinguish three regimes, which are also illustrated by snapshots of the local dissipation corresponding to the respective ranges.

Small values of the local dissipation, up to approximately $10^{-4}$, are mostly associated with the base Taylor--Green flow.
The contribution of this component to the total dissipation clearly vanishes in the high-Reynolds-number limit.
At the other extreme, large values of the local dissipation, above approximately $10^0$, are associated with very small and thin vorticity structures generated by boundary-layer detachments.
Although the local dissipation within these structures is large, their small spatial extent limits their overall contribution, as reflected by the small probability densities in this range.

Finally, intermediate values of the local dissipation, between approximately $10^{-4}$ and $10^0$, are associated with the detached vortices that populate the bulk.
These vortices make a substantial contribution to the PDFs.
However, in this range, the PDFs decrease monotonically with increasing Reynolds number, suggesting that the area occupied by the detached vortices shrinks in the high-Reynolds-number limit.

This observation is consistent with the measurements presented above, particularly with Kato's criterion for the energy dissipation rate.
Although the detached vortices become more numerous and more intense as the Reynolds number increases, their decreasing size compensates for this increase in local dissipation, resulting in a vanishing total energy dissipation in the high-Reynolds-number limit.

\subsection{Coarse grained vorticity field and fractal set of detachment points}

The mismatch of boundary conditions for the Navier--Stokes and Euler equations leads to strong vorticity concentrations near the boundary.
In the infinite-Reynolds-number limit, these concentrations may converge to a vortex sheet, corresponding to a singular vorticity distribution supported on the boundary.

\cite{kelliher2008vanishing} proved, using an argument based on Stokes' theorem, that the Navier--Stokes solutions converge to the strong Euler solution if and only if
\begin{equation}\label{EQ:convergence_vorticity}
	\omega \to \overline{\omega} - (\overline{\bu} \cdot \pmb{\tau}) \mu \qquad \text{in} \ H^1(\mathcal{D})' \quad \text{uniformly on } [0,T],
\end{equation}
where $H^1(\mathcal{D})'$ denotes the dual of the Sobolev space $H^1(\mathcal{D})$, consisting of square-integrable functions whose first weak derivatives are also square-integrable, and $\mu$ is the boundary supported measure.
For our flow configuration, this measure is given by
\begin{equation}
	\mu(\bx) = \delta(y-1/2) + \delta(y+1/2),
\end{equation}
 where $\delta$ denotes the standard one-dimensional Dirac delta distribution.
Thus, convergence in~\eqref{EQ:convergence_vorticity} means that
\begin{equation}
	\int_{\mathcal{D}} \omega(\bx,t) f(\bx) \, d\bx \to \int_{\mathcal{D}} \overline{\omega}(\bx,t) f(\bx) \, d\bx - \int_{\partial \mathcal{D}} \overline{\bu}(\bx,t) \cdot \pmb{\tau} f(\bx) dS
\end{equation}
uniformly on $t \in [0,T]$, and for every $f \in H^1(\mathcal{D})$.

This criterion characterises convergence to the Euler solution in terms of the formation of a boundary-supported vortex sheet in the limiting vorticity distribution.
In the bulk, the vorticity converges to the smooth Euler vorticity, with no additional vorticity concentrations persisting away from the boundary.
Indeed,~\cite{kelliher2008vanishing} also proved that convergence in~\eqref{EQ:convergence_vorticity} is equivalent to
\begin{equation}\label{EQ:convergence_vorticity_H_1}
	\omega \to \overline{\omega} \qquad \text{in} \ H^{-1}(\mathcal{D}) \quad \text{uniformly on } [0,T],
\end{equation}
where $H^{-1}(\mathcal{D})$ denotes the dual of $H_0^1(\mathcal{D})$, the closure of the smooth, compactly supported functions in the $H^1(\mathcal{D})$ norm.

A way to analyse vorticity concentrations is through the coarse-grained vorticity field,
\begin{equation}
	\Omega_r (\mathbf{x},t) = \frac{1}{| B_r(\bx) |}\int_{B_r(\bx)} | \omega (\bx',t)| \, d\bx'.
\end{equation}
where $B_r(\bx)$ denotes a box of side length $r$ centred at $\bx$, and $|B_r(\bx)|$ is its area.
In our numerical simulations, we discretise the domain into square boxes of side length $r$ and compute the average absolute vorticity within each box.
This yields a discrete field of coarse-grained vorticity values, $\Omega_r$, whose dependence on the coarse-graining scale $r$ we then investigate.

For a smooth vorticity field, $\Omega_r$ should scale as $r^0$ as $r\to0$.
Moreover, the maximum coarse-grained vorticity, $\max_{\bx}\Omega_r(\bx,t)$, should converge to $\sup_{\bx\in\mathcal D}|\omega(\bx,t)|$.
By contrast, if the vorticity concentrates on a line, as in the boundary-supported vortex sheet in~\eqref{EQ:convergence_vorticity}, then $\Omega_r$ should scale as $r^{-1}$.
Finally, if the vorticity contains a point vortex, represented by a Dirac delta distribution $\delta(\bx-\bx_0)$ at $\bx_0\in\mathcal D$, then $\Omega_r$ should scale as $r^{-2}$.
More generally, a non-integer scaling exponent may indicate concentration on a set with a non-integer, potentially fractal, dimension.

\begin{figure}
	\centering
	\includegraphics[width=.9\columnwidth]{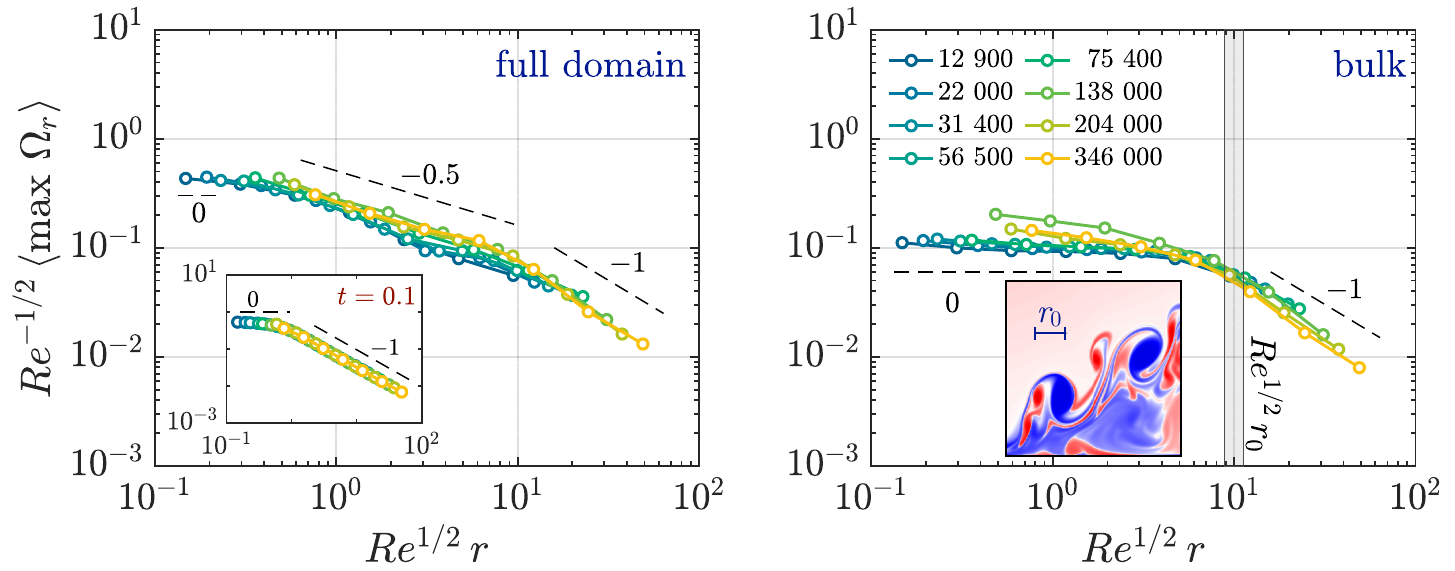}
	\caption{Time-averaged maximum coarse-grained vorticity $\langle \max \Omega_r \rangle$, rescaled by $\Rey^{-1/2}$, as a function of the coarse-grained scale $r$, rescaled by $\Rey^{1/2}$, for different Reynolds numbers. \textit{Left:} Maximum computed over the full domain, including the boundary. The inset shows the instantaneous maximum coarse-grained vorticity, $\max \Omega_r$, at time $t = 0.1$, rescaled by $\Rey^{-1/2}$, as a function of the rescaled coarse-graining scale $\Rey^{1/2}r$. \textit{Right:} Maximum computed over the bulk, excluding a region within a distance $0.0833$ of the boundary. The inset shows a vorticity snapshot for Run VIII, with $\Rey = 346\,000$, at $t = 2.9$. The reference vortex size, $r_0 = 0.0170$ is indicated in the snapshot. The corresponding rescaled scale, $\Rey^{1/2}r_0$ is highlighted in grey in the main plot. Black dashed lines indicate the corresponding reference scalings.}
	\label{fig:L1loc}
\end{figure}

Figure~\ref{fig:L1loc} shows the maximum coarse-grained vorticity, $\max_{\bx\in\mathcal D}\Omega_r$, averaged in time and rescaled by $\Rey^{-1/2}$, as a function of the coarse-graining scale $r$, rescaled by $\Rey^{1/2}$.
These rescalings reflect the Prandtl boundary-layer scaling observed for several quantities in the previous sections.
The maximum is taken over the entire domain.

We begin with the inset in Fig.~\ref{fig:L1loc}~(\textit{Left}), which shows the same quantity at the fixed time $t=0.1$.
The curves for different Reynolds numbers collapse remarkably well, reflecting the self-similar structure of the laminar boundary layers at these early times.
At large $r$, the coarse-grained vorticity scales as $r^{-1}$, consistent with the concentration of vorticity on a vortex sheet in the infinite-Reynolds-number limit.
Below a crossover scale, $\Omega_r$ instead scales as $r^0$, as expected for a smooth vorticity field.
This crossover occurs at a scale comparable to the boundary-layer thickness, which scales as $\Rey^{-1/2}$. 
The collapse of the curves in this range also indicates that the maximum vorticity scales as $\Rey^{1/2}$, consistently with the measurements in Fig.~\ref{fig:maximum_vorticity_dp_dy} for this time range.

Figure~\ref{fig:L1loc}~(\textit{Left}) also shows the measurements averaged over the turbulent regime, following the onset of boundary-layer detachment.
In this case, we observe three distinct scaling ranges.
At large $r$, $\Omega_r$ continues to scale as $r^{-1}$, consistent with the boundary-supported vortex sheet associated with convergence in~\eqref{EQ:convergence_vorticity}.
At the opposite end of the spectrum, for small $r$, we recover the scaling $\Omega_r\sim r^0$, reflecting the smoothness of the vorticity field at each finite Reynolds number.
Although not shown due to the noisy behaviour of the data, we expect the maximum coarse-grained vorticity to converge to the maximum vorticity in the domain for each Reynolds number as $r\to0$.

Interestingly, at intermediate scales, we observe a fractional power law, $\Omega_r\sim r^{-0.5}$.
This scaling is consistent with an intermediate vorticity concentration characterised by a non-integer effective dimension.
We conjecture that it may be associated with the spatial distribution of boundary-layer detachment structures along the boundary.

\begin{figure}
	\centering
	\includegraphics[width=1\columnwidth]{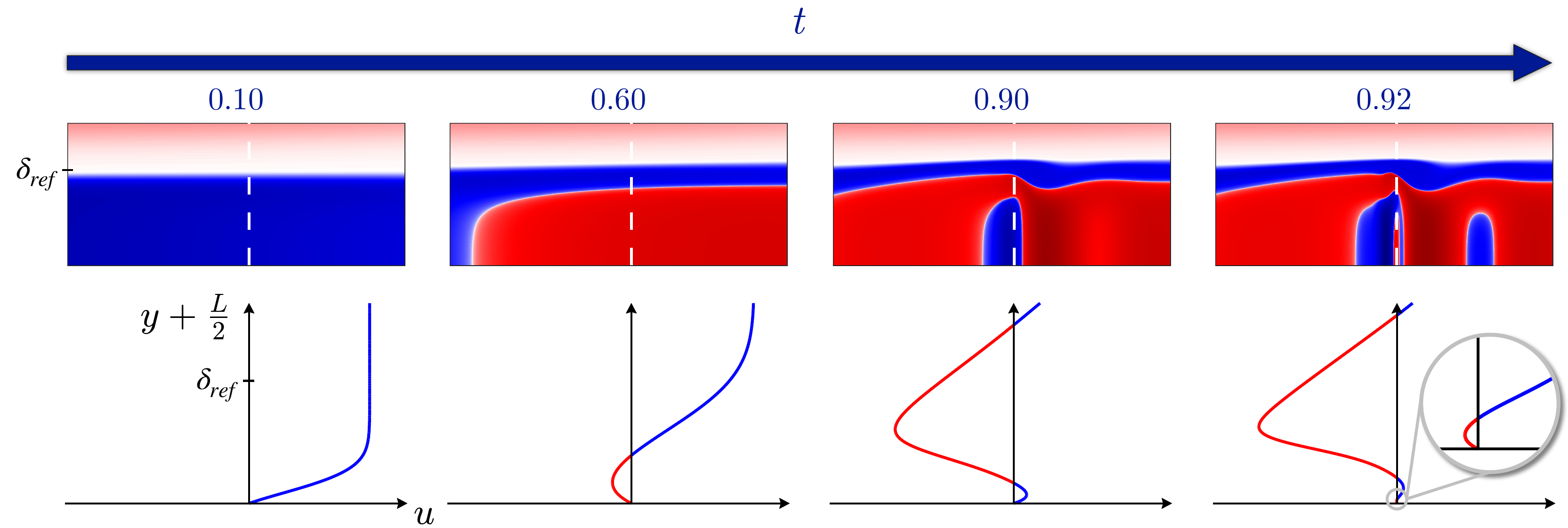}
	\caption{Close-up views of the vorticity field near the boundary at different times, together with the corresponding boundary-layer tangential velocity profiles, for Run VIII, with $\Rey = 346\,000$. The wall-normal coordinate in the snapshots is shown in logarithmic scale. The boundary-layer profiles are taken along the streamwise location indicated by the white dashed line in the corresponding snapshots. The reference wall-normal distance, $\delta_\textit{ref} = 0.005$, is indicated in the first snapshot and its corresponding boundary-layer profile.}
	\label{fig:foliation}
\end{figure}

To support this interpretation, Fig.~\ref{fig:foliation} shows vorticity snapshots in a region close to the boundary, with the wall-normal coordinate $y$ plotted on a logarithmic scale, together with the corresponding boundary-layer tangential velocity profiles at different times.
All snapshots correspond to Run VIII, with $\Rey=346\,000$.

At $t=0.10$, the boundary layer remains attached and laminar, with the characteristic negative vorticity throughout the boundary layer.
By $t=0.60$, the first signs of an adverse flow are already apparent, with positive vorticity emerging within the boundary layer.
At $t=0.90$ and $t=0.92$, we observe successive detachments, characterised by changes in the sign of the vorticity at progressively smaller distances from the wall and within increasingly narrow regions in the streamwise direction.

In particular, the snapshots reveal a foliation of detachments at the same streamwise location, where the vorticity changes sign four times as the wall-normal distance increases.
At $t=0.92$, we also observe the emergence of a second internal detachment point, initiating a new sequence of foliated detachments.
We find that the number of these foliated detachments increases with the Reynolds number.

We conjecture that this increasingly intricate arrangement of detachment points and associated vorticity structures contributes to the fractional scaling observed for $\Omega_r$ at intermediate coarse-graining scales.

Figure~\ref{fig:L1loc}~(\textit{Right}) shows the maximum coarse-grained vorticity, $\mathrm{max}_\mathcal{D} \, \Omega_r$, averaged in time over the turbulent regime and rescaled by $\Rey^{-1/2}$, as a function of the coarse-grained scale $r$, rescaled by $\Rey^{1/2}$.
In this case, the maximum is taken only over points in the bulk, at a distance of at least $0.0833$ from the boundary.

We observe two distinct scaling ranges.
At small $r$, the maximum coarse-grained vorticity scales as $r^0$, consistent with the smoothness of the flow and the absence of persistent vorticity concentrations away from the boundary.
At larger scales, however, we observe an $r^{-1}$ scaling.
Although this scaling is characteristic of a vortex-line concentration, we interpret it here as a consequence of chains of detached vortices generated at the boundary and subsequently advected into the bulk.

Indeed, the crossover scale $r_0$ between the two scaling regimes appears to scale as $\Rey^{-1/2}$, as indicated by the collapse of the curves for different Reynolds numbers.
The inset shows examples of detached vortices whose characteristic sizes are comparable to $r_0$ at the corresponding Reynolds number.
Thus, we interpret the observed $r^{-1}$ scaling as an intermediate-scale effect associated with these detached vortices, whose characteristic size decreases as the Reynolds number increases.
This scaling therefore does not indicate the formation of a persistent vortex-line concentration in the bulk. This interpretation is consistent with the convergence criterion~\eqref{EQ:convergence_vorticity_H_1}.

\section{Concluding remarks}
\label{sec:conclusions}

We have designed special initial conditions for the incompressible Navier--Stokes equations in a two-dimensional periodic channel with no-slip boundaries.
These initial conditions possess several features that make them particularly suitable for the mathematical and numerical analysis of the resulting flow.
Their formal infinite-Reynolds-number limit is an explicit, analytic, stationary solution of the incompressible Euler equations: a Taylor--Green vortex~\citep{taylor1937mechanism} with non-trivial tangential velocity and adverse pressure gradients at the boundary.

The Navier--Stokes initial data are analytic.
Consequently, the associated Prandtl boundary-layer system is locally well-posed, and the Navier--Stokes solutions converge to the Euler solution over a compact time interval~\citep{sammartino1998zeroI,sammartino1998zeroII}.
However, the boundary traces of the Euler solution also induce a well-known finite-time singularity in the Prandtl system~\citep{vandommelen1980spontaneous,kukavica2017vandommelen}.

Using highly resolved Fourier--Chebyshev spectral simulations, we computed numerical solutions of the Navier--Stokes initial-value problem at increasing Reynolds numbers.
Initially, the boundary layer remains laminar and attached to the wall.
During this regime, the maximum boundary vorticity and maximum normal pressure gradient follow the Prandtl scaling, consistent with the well-posedness of the Prandtl system and the validity of its asymptotic description.
At a finite time, however, the boundary layer detaches.
The maximum vorticity and normal pressure gradient then exhibit scalings that differ from their Prandtl predictions, suggesting a breakdown of the Prandtl system and its associated asymptotic expansion.

Repeated detachments subsequently give rise to increasingly complex, turbulent flow.
The large-scale base flow advects the detached vortices into the bulk, where they become more numerous and more intense as the Reynolds number increases.

We investigated the mechanisms of energy transfer in this turbulent state through the energy spectrum.
The spectra are consistent with a dual-cascade scenario: an inverse energy cascade at large scales, with $E\sim k^{-5/3}$, and a direct enstrophy cascade at small scales, with $E\sim k^{-\alpha}$, where $\alpha>3$.
These spectral features are not distributed uniformly across the domain.
In particular, the inverse-cascade scaling is observed only in the turbulent region near the boundary, whereas farther from the wall, the energy spectrum appears to be dominated by the advected coherent structures.
These observations are consistent with previous numerical studies in the literature, which reported similar energy-spectrum scalings~\citep{clercx2000energy,kramer2008large,hazi2010lattice}.

We employed rigorous mathematical criteria to investigate the infinite-Reynolds-number limit of the numerical solutions.
Direct measurements of the energy norm, together with numerical evaluations of Kato's criterion based on the time-averaged viscous energy dissipation~\citep{kato1984remarks} and the Bardos--Titi--Kelliher criterion based on the time-averaged power of viscous shear stresses~\citep{bardos2013mathematics,kelliher2017observations}, all yielded results consistent with convergence of the Navier--Stokes solutions to the strong Euler solution.
Remarkably, these global, time-averaged quantities continue to exhibit Prandtl-like scalings even after the breakdown of the Prandtl asymptotic description.

We further interpreted the observed flow phenomenology through a detailed statistical analysis of the vorticity distribution.
The probability density functions (PDFs) of the local energy dissipation indicate that the total area occupied by detached vortices systematically decreases.
This provides a possible explanation for the reduction in total energy dissipation, despite the fact that the detached vortices become more numerous and exhibit increasingly intense local vorticity.

The analysis of coarse-grained vorticity fields provides further evidence of the formation of a vortex sheet at the boundary and the absence of persistent vorticity concentrations in the bulk.
These properties are related to mathematical characterisations of convergence in the infinite-Reynolds-number limit~\citep{kelliher2008vanishing,kelliher2017observations}.
More strikingly, we observe an intermediate scaling regime in the coarse-grained vorticity associated with the detached vortices, suggestive of an effective fractal organisation.
This observation is consistent with the foliated structure of boundary-layer detachments revealed by vorticity snapshots: successive internal separations develop at progressively finer scales, in both the wall-normal and wall-tangential directions.

Together, our numerical findings suggest that the Navier--Stokes solutions in the flow configuration considered here converge to the strong Euler solution in the infinite-Reynolds-number limit.
To the best of our knowledge, this may provide the first numerical example of such convergence despite the breakdown of the Prandtl system and in the presence of boundary-layer detachments.
This particularly contrasts with the numerical results obtained for the dipole-wall collision, for which a Reynolds-independent energy dissipation rate is observed over the range of Reynolds numbers investigated~\citep{nguyen2011energy,nguyen2018energy}.
This, of course, raises many questions on why the two different flow configurations present such distinct behaviours in the infinite-Reynolds-number limit.
We leave this challenging and interesting study to future work.
We suggest that this question could be further investigated through comparative direct numerical simulations, focusing on detailed convergence criteria, as well as through the use of available simplified models~\citep{campolina2025logarithmic}.

The role of energy cascades and energy budgets could also be investigated further.
Rather than relying solely on one-dimensional energy spectra, as in the present work and much of the previous literature, the mechanisms of energy transfer could be characterised more directly through local energy and enstrophy fluxes.
The anisotropic nature of wall-bounded flows, however, precludes the straightforward application of standard techniques developed for homogeneous and isotropic turbulence.
In this context, the so-called filtering approach~\citep{germano1992turbulence} could provide a useful framework for examining local fluxes in different regions of the domain.
This approach has been successfully applied to investigate the locality of energy transfer in turbulent cascades~\citep{eyink1995local,xiao2009physical} and does not require the assumptions of homogeneity or isotropy~\citep{eyink2005locality}.

Following the breakdown of the Prandtl description in the flow configuration considered here, a natural question is whether more elaborate boundary-layer models could provide a more accurate description of the flow near the wall.
Possible candidates include the interactive boundary-layer (IBL) approach~\citep{carter1979new,veldman1981new,le1990new,sobey2000introduction} and the triple-deck theory~\citep{stewartson1969self,neiland1969theory,meyer1983view}.
However, these models have also recently been shown to exhibit strong ill-posedness, characterised by unrealistic high-frequency instabilities~\citep{dalibard2018high,dietert2022ill}.
Moreover, the successive internal detachments observed in our simulations raise questions about the ability of models based on a description in terms of multiple interacting layers to capture the resulting flow dynamics.
In fact, promising connections between the ill-posedness of multiple-layer models and these successive internal detachments could be investigated.

The present results could also be extended to other flow configurations.
For instance, a recent numerical study of two-dimensional flow past a cylinder reported a decrease in drag with increasing Reynolds number~\citep{chatzimanolakis2022vortex}.
The wake was also observed to become increasingly concentrated downstream of the cylinder, while the area occupied by detached vortices appeared to decrease as the Reynolds number increased.
These features suggest interesting connections with the behaviour observed in the present study and motivate further investigation of the extent to which our findings depend on the specific flow configuration.

The role of dimensionality is another important question.
Extending the present investigation to three-dimensional flows would introduce additional physical mechanisms, most notably vortex stretching, which can amplify enstrophy and alter the dynamics of the flow.
Moreover, the possible formation of finite-time singularities in the three-dimensional Euler equations would introduce further mathematical complications in analysing the inviscid limit.
These additional mechanisms make the three-dimensional problem particularly challenging and raise the question of whether convergence to the Euler solution can still be sustained in the presence of boundary-layer detachment.

%\backsection[Supplementary data]{\label{SupMat}Supplementary material and movies are available at \\https://doi.org/10.1017/jfm.2019...}

\backsection[Acknowledgements]{The authors are grateful to Theodore Drivas, Gregory Eyink, Alexei Mailybaev, Anna Mazzucato, Kai Schneider, Eric Simonnet, Simon Thalabard, and Nicolas Valade for their stimulating discussions and valuable insights.
The numerical simulations were performed using OPAL infrastructure from Universit\'{e} C\^{o}te d'Azur.}

\backsection[Funding]{This research was funded by the French National Research Agency (Projects TILT ANR-20-CE30-0035 and TURBO ANR-25-CE30-4603-01) and by the European Union (ERC NoisyFluid, No. 101053472).}

\backsection[Declaration of interests]{The authors report no conflict of interest.}

\backsection[Data availability statement]{The data supporting the findings of this study are available from the corresponding author upon reasonable request.}

\backsection[Author ORCIDs]{C. S. Campolina, https://orcid.org/0009-0000-3282-3675; J. Bec, https://orcid.org/0000-0002-3618-5743}

\backsection[Author contributions]{Both authors contributed to the design and implementation of the research, to the analysis of the results and to the writing of the manuscript.}

\appendix

\bibliographystyle{jfm}
\bibliography{references}

\end{document}